\documentclass[prd,aps,preprintnumbers,nofootinbib,superscriptaddress]{revtex4}
\usepackage{amssymb,amsmath,graphicx,bm}

\usepackage[usenames,dvipsnames]{color}
\usepackage[normalem]{ulem} 
\renewcommand\sout{\bgroup \color[rgb]{0.55,0.00,0.99} \ULdepth=-.5ex \ULset}
\newcommand{\xB}{x_{\scriptscriptstyle B}}
\newcommand{\sT}{{\scriptscriptstyle T}}

\renewcommand{\d}{\mathrm{d}}
\def\slash#1{\setbox0=\hbox{$#1$}               
        \dimen0=\wd0                            
        \setbox1=\hbox{/} \dimen1=\wd1          
        \ifdim\dimen0>\dimen1                   
        \rlap{\hbox to \dimen0{\hfil/\hfil}}    
        #1                                      
        \else
        \rlap{\hbox to \dimen1{\hfil$#1$\hfil}} 
        /                                       
        \fi}                                    %

\newcommand{\myparallel}{{\mkern3mu\vphantom{\perp}\vrule depth 0pt\mkern2mu\vrule depth 0pt\mkern3mu}}

\usepackage[normalem]{ulem} 
\renewcommand\sout{\bgroup \color[rgb]{0.55,0.00,0.99} \ULdepth=-.5ex \ULset}

\begin{document}

\title{$J/\psi$ polarization in semi-inclusive DIS at low and high transverse momentum} 

\author{Umberto D'Alesio}
\email{umberto.dalesio@ca.infn.it}
\affiliation{Dipartimento di Fisica, Universit\`a di Cagliari, Cittadella Universitaria, I-09042 Monserrato (CA), Italy}
\affiliation{INFN Sezione di Cagliari,  Cittadella Universitaria, I-09042 Monserrato (CA), Italy}

\author{Luca Maxia}
\email{luca.maxia@ca.infn.it}
\affiliation{Dipartimento di Fisica, Universit\`a di Cagliari, Cittadella Universitaria, I-09042 Monserrato (CA), Italy}
\affiliation{INFN Sezione di Cagliari,  Cittadella Universitaria, I-09042 Monserrato (CA), Italy}

\author{Francesco Murgia}
\email{francesco.murgia@ca.infn.it}
\affiliation{INFN Sezione di Cagliari,  Cittadella Universitaria, I-09042 Monserrato (CA), Italy}

\author{Cristian Pisano}
\email{cristian.pisano@unica.it}
\affiliation{Dipartimento di Fisica, Universit\`a di Cagliari, Cittadella Universitaria, I-09042 Monserrato (CA), Italy}
\affiliation{INFN Sezione di Cagliari,  Cittadella Universitaria, I-09042 Monserrato (CA), Italy}

\author{Sangem Rajesh~\footnote{Now at INFN, Sezione di Perugia, via A.~Pascoli snc, 06123, Perugia, Italy}}
\email{rajesh.sangem@pg.infn.it}
\affiliation{INFN Sezione di Cagliari,  Cittadella Universitaria, I-09042 Monserrato (CA), Italy}

\begin{abstract}
We study the polar and azimuthal decay angular distributions  of  $J/\psi$ mesons produced in semi-inclusive, deep-inelastic electron-proton scattering. For the description of  the quarkonium formation mechanism, we adopt the framework of nonrelativistic QCD, with the inclusion of the intermediate color-octet channels that are suppressed at most by a factor $v^4$ in the velocity parameter $v$ relative to the leading color-singlet channel. We put forward factorized expressions for the helicity structure functions in terms of transverse momentum dependent gluon distributions and shape functions, which are valid when the $J/\psi$  transverse momentum is small with respect to the hard scale of the process. By requiring that such expressions correctly match with the collinear factorization results at high transverse momentum,  we determine the perturbative tails of the shape
functions and find them to be independent of the $J/\psi$ polarization. In particular, we focus on the $\cos 2\phi$ azimuthal decay asymmetry, which originates from the distribution of linearly polarized gluons inside an unpolarized proton. We therefore suggest a novel experiment for the extraction of this so-far unknown parton density that could be performed,  in principle, at the future Electron-Ion Collider. 

\end{abstract}

\date{\today}

\maketitle

\section{Introduction}

The $J/\psi$ meson, {\it i.e.\ }the lightest spin-one bound state of a charm quark-antiquark pair ($c \bar c$), has been the subject of considerable attention since its simultaneous discovery at the Brookhaven National Laboratory~\cite{Aubert:1974js} and at the Stanford Linear Accelerator Center~\cite{Augustin:1974xw} in 1974. From the experimental side, it has the advantage of being abundantly produced at all high-energy electron-proton and proton-proton colliders due to its relatively low mass. Moreover, since  the $J/\psi$ meson carries the   
same total angular-momentum, parity and charge-conjugation quantum numbers of the photon,  $J^{PC} = 1^{--}$, it can decay to $e^+e^-$ and $\mu^+\mu^-$ pairs with significant branching ratios of about 6\% in both channels, thus making it especially easy to detect. From the theory side, it is an excellent  tool to probe both the perturbative and nonperturbative aspects of QCD. In fact, since the charm mass $M_c$  is much larger than the QCD scale parameter $\Lambda_\text{QCD}$, the $J/\psi$ production mechanism involves the creation of the $c\overline c$ pair, which can be treated perturbatively,  and its subsequent, nonperturbative, transition to the observed quarkonium bound state. Different frameworks provide different descriptions of the latter hadronization process. In this paper, we will adopt  the factorization approach based on the  effective field theory of nonrelativistic QCD (NRQCD)~\cite{Bodwin:1994jh}. 

According to NRQCD, the formation of a heavy-quarkonium state is described by  means of a double expansion in the strong coupling constant $\alpha_s$ and in the average relative velocity $v$ of the heavy quark-antiquark pair in the quarkonium rest frame, with  $v^2 \simeq 0.3$ for charmonium and $v^2 \simeq 0.1$ for bottomonium.  The main feature of this approach is that, at short distances, a heavy quark-antiquark pair can be produced not only directly in a color-singlet (CS) configuration, but also in a color-octet (CO) state, which subsequently evolves into the observed quarkonium through soft gluon radiation. Such nonperturbative hadronization process is encoded in the so-called long-distance matrix elements (LDMEs), which are expected to be universal. The CS LDMEs are well known from potential models, lattice calculations and leptonic decays.  Conversely, the CO ones are typically extracted from fits to data on $J/\psi$ and $\Upsilon$ yields, see for instance Refs.~\cite{Butenschoen:2010rq,Chao:2012iv,Sharma:2012dy,Bodwin:2014gia,Zhang:2014ybe}, but not yet from lattice calculations.  

Although NRQCD successfully explains many experimental observations, it still presents major challenges~\cite{Brambilla:2010cs,Andronic:2015wma,Lansberg:2019adr}. On the one hand, 
the present knowledge of the CO matrix elements is not very accurate, because the different sets of their extracted values are not compatible with each other, even within the large uncertainties. For this reason, new ways to determine them with better precision have been recently put forward~\cite{Qiu:2020xum,Boer:2021ehu}. On the other hand, phenomenological analyses within the NRQCD framework are not able to consistently account for  all cross sections and polarization measurements for $J/\psi$ mesons produced both in proton-proton and in electron-proton collisions. 

In this paper, we show how the study of $J/\psi$ polarization in semi-inclusive, deep-inelastic electron-proton scattering (SIDIS), {\it i.e.}\ $e\, p \to e^\prime\, J/\psi\, X$,  can shed light on the still open puzzles of quarkonium production mechanism and polarization. Such an analysis is complementary to the one performed in Ref.~\cite{Boer:2020bbd}, where the transverse momentum spectrum and the azimuthal distribution for the production of unpolarized $J/\psi$ mesons  were considered. A satisfactory theoretical description will be achieved only if all these observables can be simultaneously described within a unique framework. Their experimental  determination could be reached at  the future Electron-Ion Collider (EIC)~\cite{AbdulKhalek:2021gbh,Accardi:2012qut,Boer:2011fh}, which will be built in the United States.

Along the lines of Ref.~\cite{Boer:2020bbd}, in order to avoid contributions from photoproduction processes, we only consider the kinematic region where the virtuality of the photon exchanged in the reaction, $Q$, is equal or greater than the $J/\psi$ mass, $M_\psi$, namely $Q\ge M_\psi$. Moreover, we denote by $q_\sT$ the transverse momentum of the photon with respect to the $J/\psi$ and the proton four-momenta. Hence, in the calculation of the polarization parameters, we have to deal with  three relevant scales: the above-defined transverse momentum $q_\sT$, a hard scale $\mu$ (to be identified with $Q$, or $M_\psi$, or any combination of them) and a soft scale (the nonperturbative QCD scale $\Lambda_{\text{QCD}}$ or, alternatively, the proton mass). Depending on the value of $q_\sT$ we can adopt two different factorization frameworks, both of them allowing for a separation of the short-distance from the long-distance contributions to the  observables under study. 

In the high-$q_\sT$ region, namely $q_\sT \gg \Lambda_{\rm QCD}$, in a  frame where the $J/\psi$ meson is at rest, the photon transverse momentum is generated by perturbative radiation. NRQCD and collinear factorization can be applied and the resulting polarization parameters will depend on collinear ({\it i.e.}\ integrated over transverse momentum) parton distribution functions (PDFs) and LDMEs. 
In the small-$q_\sT$ region, $q_\sT\ll \mu$, the photon transverse momentum is  nonperturbative instead, and transverse momentum dependent (TMD) factorization~\cite{Collins:2011zzd,GarciaEchevarria:2011rb,Echevarria:2012js} is expected to be appropriate. Observables should then depend on TMD PDFs (or TMDs for short) and shape functions~\cite{Echevarria:2019ynx,Fleming:2019pzj}, which are the generalization of the NRQCD LDMEs. Alternatively, the latter can be seen as the analog of the TMD fragmentation functions for light hadrons. We note that in the  overlapping region, $\Lambda_{\text{QCD}} \ll q_\sT \ll \mu$, both frameworks can be applied and the results obtained within the two formalisms have to match, provided they describe the same underlying mechanism~\cite{Bacchetta:2008xw}. This property has been proven for several observables for which TMD factorization at the twist-two level has been demonstrated, such as 
the unpolarized cross sections, differential in $q_\sT$ and integrated over the azimuthal angles of the final particles,  for the SIDIS process $e\,p\to e^\prime\, h\, X$, where $h$ is a light hadron, and for Drell-Yan (DY) dilepton production,  $p\, p\to \ell\, \ell^\prime\,X$~\cite{Collins:1984kg,Catani:2000vq}.  

Although there is not yet a rigorous proof of TMD factorization for the  $J/\psi$  polarization parameters in SIDIS, 
strong arguments exist in favor of its validity, if we consider the analogy with $e\,p \to e^\prime \, h \,X$, for which TMD factorization holds at all orders~\cite{Collins:2011zzd}. As already pointed out in Ref.~\cite{Boer:2020bbd},  these two processes are essentially equivalent from the point of view of the color flow, which determines the gauge-link structure of the TMD parton correlators~\cite{Bacchetta:2018ivt}. Since neither the $J/\psi$ mass nor its spin can affect such structure, we do not expect any factorization breaking effects due to color entanglement. We therefore propose reasonable factorized expressions for the $J/\psi$ polarization parameters in terms of twist-two TMDs and shape functions, which properly match with the collinear results in the intermediate region $\Lambda_{\text{QCD}} \ll q_\sT \ll \mu$. 

Finally, we point out that the theoretical TMD framework we have devised will also have important implications for the extraction of gluon TMDs at the EIC.  Indeed, although several proposals for the extraction of these distributions have been put forward,  both in $ep$~\cite{Yuan:2008vn,Bacchetta:2018ivt,Mukherjee:2016qxa,DAlesio:2019qpk} and in $pp$ collisions~\cite{Yuan:2008vn,Dunnen:2014eta,Lansberg:2017tlc,Lansberg:2017dzg,Scarpa:2019fol}, as well as within the more phenomenological generalized parton model approach~\cite{Godbole:2013bca,Godbole:2014tha,Godbole:2017syo,Kishore:2018ugo,Rajesh:2018qks,DAlesio:2017rzj,DAlesio:2018rnv,DAlesio:2019gnu}, they  are still basically unkown. In particular, we find that the so-called $\nu$ parameter, related to the $\cos 2\phi$ azimuthal asymmetry of the leptons from the $J/\psi$ decay, can give direct access to the distribution of linearly polarized gluons inside unpolarized protons.  

The paper is organized as follows. In Section~\ref{sec:structure-f} we review in detail the main properties of the cross section using only kinematic considerations and the symmetries of the strong and electromagnetic interactions, without referring to any specific model concerning  quarkonium formation. In particular, the well-known result of the angular structure of the cross section, expressed in terms of four independent helicity structure functions, is obtained along the lines of the derivation presented in Ref.~\cite{Lam:1978pu} for the DY process.  In Section~\ref{sec:coll} we compute the polarized structure functions within the framework of collinear factorization and NRQCD at the order $\alpha_s^2$. Moreover, we investigate  their small-$q_\sT$ limit and show that the dominant terms do not depend  on the choice of the coordinate axes. Section~\ref{sec:TMD} is devoted  to the calculation of the structure functions in the TMD regime at the order $\alpha_s$ and to the study of their matching  with the collinear results in the common region of validity.  Summary and conclusions are gathered in Section~\ref{sec:conc}. Details on the reference frames and on the transformations which connect them can be found in Appendix~\ref{sec:frames}. Finally, the explicit expressions for the partonic helicity structure functions in the Gottfried-Jackson frame are collected in Appendix~\ref{sec:helicity}.

\section{Model independent  properties of the cross section}
\label{sec:structure-f}

We study the process 
\begin{equation}
e(k) + p(P) \to e(k^\prime) + J/\psi(P_\psi) + X(P_X) \,,
\label{eq:reaction}
\end{equation}
with $X$ being an undetected hadronic system,
and the subsequent leptonic decay 
\begin{align}
J/\psi (P_\psi)\to \ell^+(l)\, + \,  \ell^-(l^\prime)\,,
\end{align}
where the four-momenta of the particles are given within brackets.  Throughout this work the mass of the leptons, both in the initial and in the final state,  are neglected, while we denote by $M_p$ and $M_\psi$ the proton and the $J/\psi$ masses, respectively.  Moreover, we will sum over the lepton and $J/\psi$ helicities. The virtual photon exchanged in the reaction carries four-momentum 
\begin{align}
q =  k-k^\prime\,,
\end{align}
with $Q^2 \equiv -q^2>0$. In the deep-inelastic limit, where $Q^2$, $P\cdot q$ and $P_\psi\cdot q$ are large while the variables
\begin{equation}
\xB = \frac{Q^2}{2P\cdot q}\,, \qquad y = \frac{P\cdot q}{P\cdot k}\, , \qquad z = \frac{P\cdot P_\psi}{P\cdot q}
\label{eq:SIDIS-var}
\end{equation}
are finite (with values between zero and one), the square of the invariant amplitude for this process can be split into a leptonic part and a purely hadronic part,  
\begin{align}
\vert {\cal M}\vert^2  = \frac{1}{Q^4}\, L^{\mu \nu}H_{\mu\nu}\,,
\end{align}
where the leptonic tensor $L^{\mu\nu}$ reads
\begin{align}
L^{\mu\nu} & = e^2 \left [- g^{\mu\nu} Q^2 + 2 ({k}^{\mu}{k}^{\prime \nu} + {k}^\nu {k}^{\prime \mu}) \right ]  \,,
\label{eq:lep-tensor}
\end{align}
and $H_{\mu\nu}$ is the product of the hadronic current matrix elements,
\begin{align}
H_{\mu\nu} = \langle P \vert J_\mu(0)\vert P_X; P_\psi\rangle \langle P_X; P_\psi \vert J_\nu(0)\vert P \rangle \,.
\label{eq:hadr-current}
\end{align}
The corresponding cross section reads
\begin{align}
\d \sigma = \frac{1}{2S}\, \vert {\cal M}\vert^2\, B_{\ell \ell }\,(2\pi)^4 \delta^4 (q+P- l - l^\prime- P_X)\,\frac{\d^3 P_X}{(2\pi)^3\,2 P^0_X} \,\frac{\d^3 k'}{(2\pi)^3\,2 E_e^{\prime}}\,  \frac{\d^3 l}{(2\pi)^32 l^0 }\, \frac{\d^3 l^\prime}{(2\pi)^32 l^{\prime 0} }\,,
\label{eq:cs-0}
\end{align}
where $S =  (k + P)^2 \approx 2\,k\cdot P$ if also the proton mass is neglected, while $B_{\ell\ell}$ is the branching ratio for the decay process $J/\psi \to \ell^+ \ell^-$. The integration of $H_{\mu\nu}$ over $P_X$ leads to the usual hadronic tensor 
\begin{align}
W_{\mu\nu} = 
\int \!\frac{\d^3 P_X}{(2\pi)^3\,2 P^0_X} \, \delta^4 (q+P - P_\psi-P_X)\, H_{\mu\nu}\,.
\end{align}
Moreover, it is convenient to express the phase space of the final leptons as follows
\begin{align}
\frac{\d^3 k^\prime}{(2\pi)^32E^\prime_e}  & =\frac{1}{4(2\pi)^3}\, S y \,\d \xB \,\d y\, \d \phi_{k^\prime}, \nonumber\\
 \frac{\d^3 l}{(2\pi)^32 l^0 }\, \frac{\d^3 l^\prime}{(2\pi)^32 l^{\prime 0} }
& =  \frac{\d^4P_\psi}{(2 \pi)^4}\, \frac{\d\Omega}{32\pi^2 }\,,
\end{align}
where $\phi_{k^\prime}$ is the azimuthal angle of the scattered electron, while $\Omega = (\theta, \varphi)$ refers to the solid angle of lepton $\ell^+$ in a reference frame where the system formed by $\ell^+$ and $\ell^-$ is at rest. Hence, after integration over $\phi_{k^\prime}$, Eq.~\eqref{eq:cs-0} can be cast in the form
\begin{align}
\frac{1}{B_{\ell \ell}}\, \frac{\d\sigma}{\d \xB\,  \d y\,\d^4 P_\psi\,\d \Omega } = \frac{1}{4 (4 \pi)^4}\, \frac{y}{Q^4}\, L^{\mu\nu}W_{\mu\nu}\, \,.
\label{eq:cs-LW}
\end{align}

The hadronic tensor $W_{\mu\nu}$ is a function of the four-vectors $q^\mu$, $P^\mu$, $P_\psi^\mu$ and contains the information on the proton structure as well as on the $J/\psi$ formation process and polarization. It has to fulfill certain constraints due to electromagnetic gauge invariance, parity, and hermiticity. More specifically,  the gauge-invariance condition,
\begin{align}
q^\mu W_{\mu\nu}(q,P,P_\psi) & = q^\nu W_{\mu\nu}(q,P,P_\psi)  = 0\,, 
\label{eq:gauge}
\end{align}
is implied by conservation of the hadronic current $J_\mu(0)$ in Eq.~\eqref{eq:hadr-current} and  limits $W_{\mu\nu}$  to be a tensor in the three-dimensional space orthogonal to $q^\mu$.  
Moreover, as a consequence of the hermiticity condition,
\begin{align}
W_{\mu \nu}(q,P,P_\psi) & = W^*_{\nu \mu}(q,P,P_\psi) \,,
\label{eq:hermit}
\end{align}
the symmetric part of the hadronic tensor in the indices $(\mu,\nu)$ is real, while its antisymmetric part is imaginary. Since the leptonic tensor in Eq.~\eqref{eq:lep-tensor} is symmetric, only the symmetric part of $W_{\mu\nu}$ contributes to the cross section. Finally, parity conservation implies
\begin{align}
W_{\mu\nu}(q,P,P_\psi) = W_{\mu\nu}(\overline q,\overline P,\overline P_\psi) \,,
\label{eq:parity}
\end{align}
with the definition $\overline v^\mu \equiv v_\mu = (v^0, -\bm v)$ for any vector $v^\mu=(v^0, \bm v)$. By imposing Eqs.~\eqref{eq:gauge} and \eqref{eq:parity}, the hadronic tensor can be decomposed into a set of  four basis tensors multiplied by scalar functions, the so-called {\em structure functions}, which have to be real because of Eq.~\eqref{eq:hermit}. To this end, it is useful to introduce  the projector operator to the space orthogonal to $q^\mu$, 
\begin{align}
\widehat g^{\mu \nu} = g^{\mu\nu} + \frac{q^\mu q^\nu}{Q^2}\,,
\end{align}
satisfying the relation  $\widehat g_{\mu\nu} q^\mu = \widehat g_{\mu\nu} q^\nu = 0$ and such that, when contracted with any four-vector, yields a vector orthogonal to $q^\mu$.  In particular, 
\begin{align}
\widehat P^\mu  =\frac{1}{M_p} \,\widehat g^{\mu\nu} P_\nu = \frac{1}{M_p}\,\left ( P^\mu + \frac{P\cdot q}{Q^2} \,q^\mu  \right ) \,, \qquad 
\widehat P_\psi^\mu = \frac{1}{M_\psi} \,\widehat g^{\mu\nu} P_{\psi \nu} = \frac{1}{M_\psi}\,\left ( P_\psi^\mu + \frac{P_\psi \cdot q}{Q^2} \,q^\mu  \right ) \,,
\end{align}
with $\widehat P\cdot q = \widehat P_\psi\cdot q=0$.  Hence, the hadronic tensor can be expressed as 
\begin{align}
W^{\mu\nu}(q,P,P_\psi) = - W_1\,  \widehat g^{\mu\nu} \, +\, W_2\, \widehat{P}^\mu \widehat{P}^\nu 
-\frac{1}{2}\, W_3 \,(\widehat{P}^\mu \widehat{P}_\psi^\nu+\widehat{P}_\psi^\mu \widehat{P}^\nu) + W_4\, \widehat{P}_\psi^\mu \widehat{P}_\psi^\nu \,,
\label{eq:had-tensor}
\end{align}
in terms of the common Lorentz invariant structure functions $W_{i}=W_{i}(Q^2, q\cdot P, q\cdot P_\psi)$, with $i=1,2,3,4$. 

An alternative parametrization of the hadronic tensor is provided by four, frame dependent, helicity structure functions, which, as already pointed out in Ref.~\cite{Lam:1978pu} for the DY process, have the advantage of allowing for an explicit  factorization of the leptonic and hadronic variables in the cross section. In the following, we will show that this is the case also for the process under study.

As a first step of our derivation, it might be useful to  write separately the contributions coming from transversely  and  longitudinally polarized virtual photons. This can be achieved  by expressing the leptonic tensor in Eq.~\eqref{eq:lep-tensor} in the form
\begin{align}
 L^{\mu\nu} =e^2\,  \frac{Q^2}{y^2}\, \bigg  \{ -[ 1+(1-y)^2 ]\, g_\perp^{\mu\nu} \,+ \, 4 (1-y)\,  \epsilon^\mu_{\gamma \myparallel} \epsilon^\nu_{\gamma \myparallel}  \bigg \} \,,
\label{eq:lep-2}
\end{align}
which is valid upon integration over the azimuthal angle of the electron scattering plane, taken with respect to the plane formed by the proton and the virtual photon, and neglecting the proton mass. In Eq.~\eqref{eq:lep-2}, we have introduced the projector to the space orthogonal to both $q$ and $P$ 
\begin{align}
g_\perp^{\mu\nu} & 
= g^{\mu\nu} - \frac{1}{P\cdot q} \, (P^\mu q^\nu + P^\nu q^\mu) - \frac{Q^2}{(P\cdot q)^2}\,P^\mu P^\nu
\label{eq:gperp}
\end{align}
and the  longitudinal polarization vector of the exchanged photon, 
\begin{align}
\epsilon_{\gamma { \myparallel}}^\mu (q)= \frac{1}{Q}\left (  q^\mu + \frac{Q^2}{P\cdot q} \, P^\mu \right ) \,,
\label{eq:epsL-gamma}
\end{align}
which fulfills the relations $\epsilon^2_{\gamma \myparallel} (q)= 1$ and $\epsilon_{\gamma \myparallel}^\mu(q)\, q_\mu = 0$. Hence we obtain
\begin{align}
L^{\mu\nu} W_{\mu\nu}= e^2 \, \frac{Q^2}{y^2}\,\left  \{ [1+(1-y)^2] \, {\cal W}^{\perp} \,+ \, (1-y )\,  {\cal W}^{\myparallel}  \right \}\,,
\label{eq:LW}
\end{align}
with the definitions
\begin{align}
{\cal W}^{\perp}  \equiv - g_\perp^{\mu\nu}  W_{\mu\nu} = - {W^\mu}_\mu + \frac{Q^2}{( P\cdot q)^2}\,  P^\mu  P^\nu\, W_{\mu\nu} \,, \qquad \qquad
{\cal W}^{\myparallel}  \equiv  4\, \epsilon^\mu_{\gamma \myparallel} \epsilon^\nu_{\gamma \myparallel} W_{\mu\nu}  = \frac{4 \,Q^2}{( P\cdot q)^2}\,  P^\mu P^\nu\, W_{\mu\nu}\,,
\end{align}
which in turn can be expressed in terms of either the invariant structure functions or the helicity ones. 
\begin{figure}[t]
\begin{centering}
\includegraphics[width =2.2in]{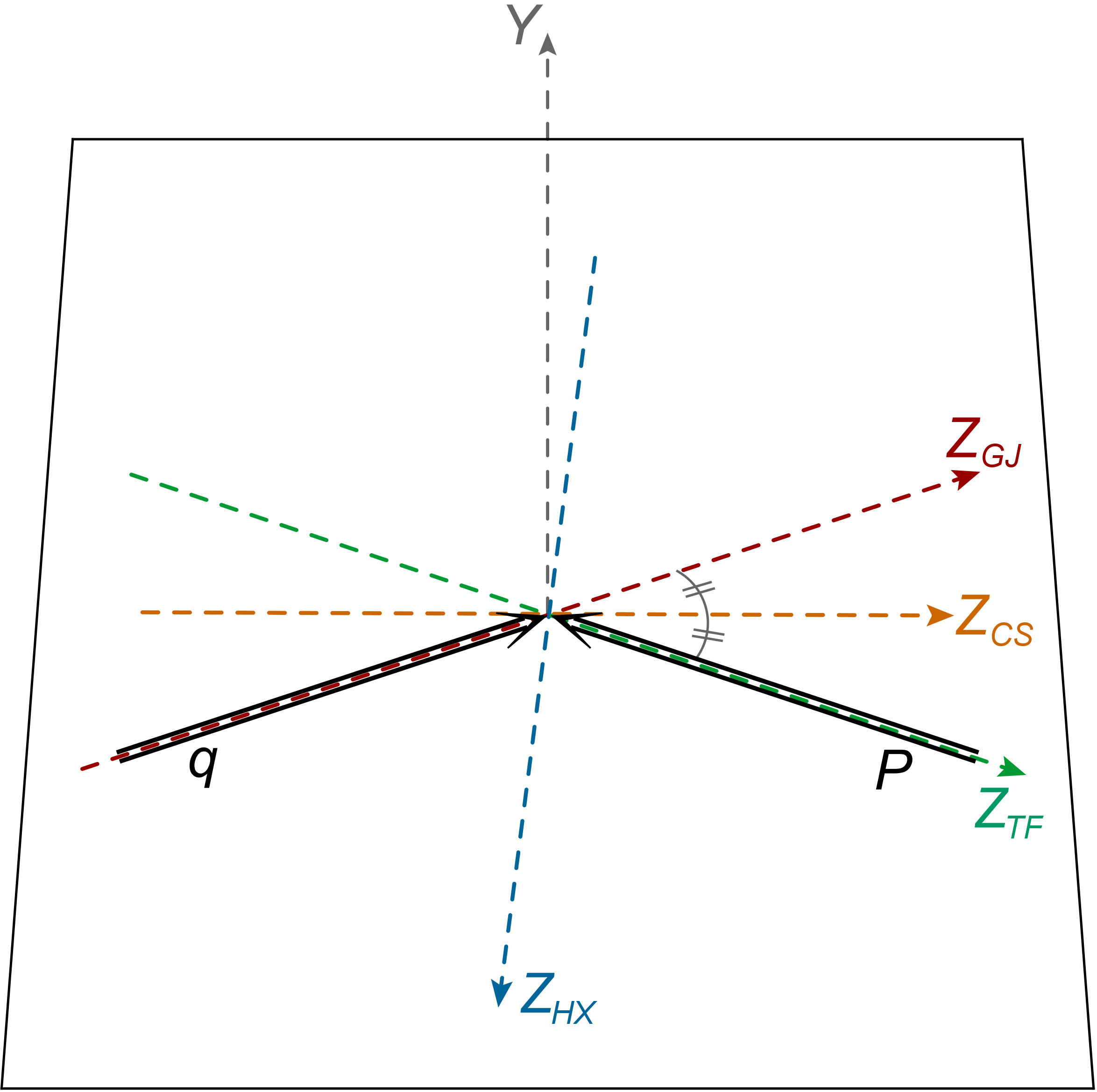}
\end{centering}
\caption{Reference frames for the process $\gamma^*(q) + p(P)\to J/\psi (P_\psi) +  X$ in which the $J/\psi$ is at rest: Helicity (HX), Target (TF), Collins-Soper (CS) and Gottfried-Jackson (GJ). They are related to each other by a rotation around the $Y$-axis. See Appendix~\ref{sec:frames} for further details.}
\label{fig:frames}
\end{figure}

It still remains to clarify the dependence of the cross section on the angles of the decaying lepton $\ell^+$.  To this end, we write the sum over  the final $J/\psi$ helicities $\lambda, \lambda^\prime = -1,0,1$ in an explicit form,
\begin{align}
{\cal W}^{\cal P} = \sum_{\lambda,\lambda^\prime} {\cal W}^{\cal P}_{\alpha\beta}\, \epsilon^\alpha_\lambda(P_\psi)\, \epsilon^{\beta\,*}_{\lambda^\prime}(P_\psi)\, \delta_{\lambda\lambda^\prime} = \sum_{\lambda, \lambda^\prime} {\cal W}^{\cal P}_{\lambda\lambda^\prime}\, \delta_{\lambda\lambda^\prime}\,, 
\label{eq:WP}
\end{align}
where the superscript ${\cal P} = \perp, \myparallel$ refers to the photon polarization and we have introduced the frame-dependent helicity structure functions
\begin{align}
{\cal W}^{\cal P}_{\lambda \lambda^\prime}\equiv\epsilon^\alpha_\lambda(P_\psi)\,\epsilon^{\ast\beta}_{\lambda^\prime}(P_\psi)\,{\cal W}_{\alpha\beta}^{\cal P}\,,
\end{align}
with $\epsilon^\mu_\lambda(P_\psi)$ being the polarization vectors of the spin-1 $J/\psi$ meson defined with respect to a covariant set of coordinate axes $(T^\mu,X^\mu,Y^\mu,Z^\mu)$  with normalizations $T^2=1 $ and  $X^2=Y^2=Z^2=-1$. These four-vectors can be defined as linear combinations of the physical momenta $q^\mu$, $P^\mu$, $P_\psi^\mu$, in such a way that $X^\mu$, $Y^\mu$, $Z^\mu$ become three-vectors in the quarkonium rest frame~\cite{Lam:1978pu,Beneke:1998re}. An illustration of the commonly used frames is given in Fig.~\ref{fig:frames}. In all of them, 
\begin{align}
T^\mu = \frac{P_\psi^\mu}{M_\psi}\,, \qquad \qquad Y^\mu = \varepsilon^{\mu\nu\alpha\beta} X_{\nu}Z_\alpha T_\beta\,,
\label{eq:T-Y}
\end{align}
with $\varepsilon^{0123} =+1$, while they differ for the choice of $X^\mu$ and $Z^\mu$, see  also Appendix~\ref{sec:frames} for details. Once the frame is fixed, the polarization vectors are given by
\begin{align}
\epsilon^\mu_0({\bm P_\psi})=Z^\mu=(0,\,0,\,0,\,1)\,,\qquad \epsilon^\mu_{\pm1}(\bm P_\psi)=\frac{1}{\sqrt{2}}(\mp X^\mu -i Y^\mu) =\frac{1}{\sqrt{2}}\, (0, \mp 1, -i ,0) \,,
\end{align}
and fulfill the orthogonality and completeness relations, 
\begin{align}
\epsilon_{\lambda\, \alpha}^*(\bm P_\psi) \, \epsilon^\alpha_{\lambda^\prime} (\bm P_\psi)& = -\delta_{\lambda\lambda^\prime}\,, \nonumber \\
\sum_{\lambda = -1,0,1}\, \epsilon_\lambda^{\alpha*} (\bm P_\psi)\, \epsilon^\beta_{\lambda} (\bm P_\psi)& = -g^{\alpha\beta} \,+ \,\frac{P^\alpha_\psi P^\beta_\psi}{M_\psi^2}\,.
\label{eq:pol-vec-2}
\end{align}
The constraints of parity conservation and hermiticity imposed by QCD on the hadronic tensor imply the following relations for the helicity structure functions
\begin{align}
{\cal W}_{\lambda\lambda^\prime}^{\cal P}  = {\cal W}_{\lambda^\prime\lambda}^{\cal P\,*} \,, \qquad \qquad 
{\cal W}_{\lambda\lambda^\prime}^{\cal P} = (-1)^{\lambda + \lambda^\prime}{\cal W}_{-\lambda -\lambda^\prime}^{\cal P} \,.
\end{align}
We can therefore  decompose ${\cal W}_{\alpha\beta}^{\myparallel}$ and ${\cal W}_{\alpha\beta}^{\perp}$ as \cite{Lam:1978pu,Boer:2006eq}
\begin{align}
{\cal W}^{\cal P}_{\alpha\beta}= -({\cal W}^{\cal P}_T + {\cal W}^{\cal P}_{\Delta\Delta})\,  ({g}_{\alpha\beta}-T_\alpha T_\beta) + ({\cal W}^{\cal P}_L-{\cal W}^{\cal P}_T - {\cal W}^{\cal P}_{\Delta\Delta}) \, Z_\alpha Z_\beta \,-\, {\cal W}^{\cal P}_\Delta\, (X_\alpha Z_{\beta} +  Z_{\alpha}  X_\beta) \, - \,2 {\cal W}^{\cal P}_{\Delta\Delta} \, X_{\alpha} X_{\beta} \,,
\label{eq:had-2}
\end{align}
in terms of the eight independent helicity structure functions
\begin{align}
{\cal W}^{\cal P}_T  & \equiv {\cal W}^{\cal P}_{1 1} = {\cal W}^{\cal P}_{-1-1}\,,\nonumber \\
{\cal W}^{\cal P}_L & \equiv {\cal W}^{\cal P}_{0 0}   \,,\nonumber \\
{\cal W}^{\cal P}_\Delta & \equiv \frac{1}{\sqrt{2}}\, ({\cal W}^{\cal P}_{1 0} + {\cal W}^{\cal P}_{0 1}) =  \sqrt{2}\, \text{Re}\, {\cal W}^{\cal P}_{ 1 0}\,,\nonumber \\
{\cal W}^{\cal P}_{\Delta \Delta} & \equiv {\cal W}^{\cal P}_{1 -1} ={\cal W}^{\cal P}_{-1 1} \,,
\label{eq:W-P-L}
\end{align}
where the subscripts refer to the $J/\psi$ polarization: ${\cal W}^{\cal P}_T$ and ${\cal W}^{\cal P}_L$ are respectively the structure functions for transversely and longitudinally polarized $J/\psi$ mesons,  ${\cal W}^{\cal P}_\Delta$  are the single-helicity flip structure functions, while ${\cal W}^{\cal P}_{\Delta\Delta}$ are the double-helicity flip ones. 

A further simplification occurs because the leptonic decay conserves helicity in the massless limit we are considering,
 {\it i.e.}\ the lepton spins do not flip in the coupling with the $J/\psi$ meson. In the quarkonium rest frame, where $\ell^+$ and $\ell^-$ are produced back-to-back along the direction identified by the three-vector 
\begin{align}
\bm {l} = (\sin\theta\cos\varphi, \sin\theta\sin\varphi, \cos\theta)\,,
\end{align}
the component along this direction of the total angular momentum of the dilepton system, and therefore of the $J/\psi$, can be either $+1$ or $-1$, but not zero. Hence, the sum in Eq.~\eqref{eq:WP} can be performed  by choosing a specific set of polarization vectors $\epsilon^\alpha_\sigma(P_\psi)$, where $\sigma$ is the $J/\psi$ helicity along the direction ${\bm l}$,
\begin{align}
{\cal W}^{\cal P} = {\cal W}^{\cal P}_{\alpha\beta}\, \sum_{\sigma= -1, 1}  \epsilon^\alpha_\sigma(P_\psi)\, \epsilon^{\beta\,*}_{\sigma}(P_\psi) = - {\cal W}^{\cal P}_{\alpha\beta} \, g_{l \perp}^{\alpha\beta} \,,
\label{eq:WP2}
\end{align}
and  the transverse projector with respect to the directions of  the four-vectors $P_\psi$ and $l$ is given by
\begin{align}
 g_{l \perp}^{\alpha\beta}  & = g^{\alpha\beta} -\frac{1}{l\cdot P_\psi} \, (l^\alpha P_\psi^\beta \,+ \,l^\beta P_\psi^\alpha) + \frac{M_\psi^2}{(l\cdot P_\psi)^2}\,l^\alpha l^\beta\,.
\end{align}
By contracting the above expression for $g_{l \perp}^{\alpha\beta}$  with the one for ${\cal W}^{\cal P}_{\alpha\beta}$ in Eq.~\eqref{eq:had-2}, according to Eq.~\eqref{eq:WP2} we obtain
\begin{align}
{\cal W}^{\cal P} = {\cal W}^{\cal P}_T  (1+\cos^2\theta)\,+\, {\cal W}^{\cal P}_L (1-\cos^2\theta)\,+\, {\cal W}^{\cal P}_\Delta \sin 2\theta \cos\varphi \, +\,  {\cal W}^{\cal P}_{\Delta\Delta}  \sin^2\theta \cos 2\varphi \,.
\end{align}
Hence, by using this result and the one in Eq.~\eqref{eq:LW}, the angular structure of the cross section in Eq.~\eqref{eq:cs-LW} can be made explicit. We find
\begin{align}
\frac{1}{B_{\ell \ell}}\, \frac{\d\sigma}{\d \xB\,  \d y\,\d^4 P_\psi\,\d \Omega } = \frac{1}{4(4 \pi)^3} \,\frac{\alpha}{y Q^2}\, \left [{\cal W}_T  (1+\cos^2\theta)\,+\, {\cal W}_L (1-\cos^2\theta)\,+\, {\cal W}_\Delta \sin 2\theta \cos\varphi \, +\,  {\cal W}_{\Delta\Delta}  \sin^2\theta \cos 2\varphi \right ]\,,
\label{eq:cs-angles}
\end{align}
with the definitions
\begin{align}
{\cal W}_{\Lambda} \equiv  \left [  1+ (1-y)^2\right ]  {\cal W}^{ \perp}_{\Lambda}  \,+ \,  (1-y) {\cal W}^{ \myparallel}_{\Lambda} \, \qquad \text{with} \qquad \Lambda = T, L , \Delta, \Delta\Delta\,.
\label{eq:W-Lambda}
\end{align}
We point out that Eq.~\eqref{eq:cs-angles} represents the main result of this section and shows, as anticipated above, that the helicity structure functions  defined in Eqs.~\eqref{eq:W-P-L} and \eqref{eq:W-Lambda} allow for  an explicit factorization of the leptonic and hadronic variables in the cross section. 

It is also convenient to  introduce the following ratio of differential cross sections~\cite{Boer:2006eq}, 
\begin{align}
\frac{\d N}{\d \Omega} \equiv \left ( \frac{\d\sigma}{\d \xB\,  \d y\,\d^4 P_\psi}  \right )^{-1}\frac{\d\sigma}{\d \xB\,  \d y\,\d^4 P_\psi\,\d \Omega } \,,
\end{align}
which is therefore given by
\begin{align}
\frac{\d N}{\d \Omega} = \frac{3}{8 \pi}\, \frac{{\cal W}_T  (1+\cos^2\theta)\,+\, {\cal W}_L (1-\cos^2\theta)\,+\, {\cal W}_\Delta \sin 2\theta \cos\varphi \, +\,  {\cal W}_{\Delta\Delta}  \sin^2\theta \cos 2\varphi }{2{\cal W}_T + {\cal W}_L}\,.
\label{eq:N-angles}
\end{align}
Alternatively, this ratio can be expressed in terms of the largely used polarization parameters $\lambda$, $\mu$, $\nu$,
\begin{align}
\frac{\d N}{\d \Omega} = \frac{3}{4 \pi}\, \frac{1}{\lambda + 3}\, \left [1 + \lambda \cos^2\theta + \mu  \sin 2\theta \cos\varphi  +  \frac{1}{2} \, \nu\,   \sin^2\theta \cos 2\varphi \right ]\,,
\label{eq:N-angles-2}
\end{align}
where, by comparison with Eq.~\eqref{eq:N-angles}, we have 
\begin{align}
\lambda = \frac{{\cal W}_T -{\cal  W}_L}{{\cal W}_T + {\cal W}_L}\,, \qquad \mu = \frac{{\cal W}_\Delta}{{\cal W}_T + {\cal W}_L}\, , \qquad \nu = \frac{2 {\cal W}_{\Delta \Delta}}{{\cal W}_T + {\cal W}_L}\,.
\end{align}
Notice that Eqs.~\eqref{eq:N-angles} and \eqref{eq:N-angles-2} describe the typical angular distribution of the leptons that originate from the decay of a spin-one particle in its rest frame, for instance a virtual photon produced in the Drell-Yan process~\cite{Boer:2011fh} or a $J/\psi$ meson produced in photon-proton collisions~\cite{Beneke:1998re}. 

\section{Helicity structure functions within NRQCD and collinear factorization}
\label{sec:coll}

\begin{figure}[t]
\begin{centering}
\includegraphics[trim={0cm 0.cm 0cm 0cm},clip,scale=0.5]{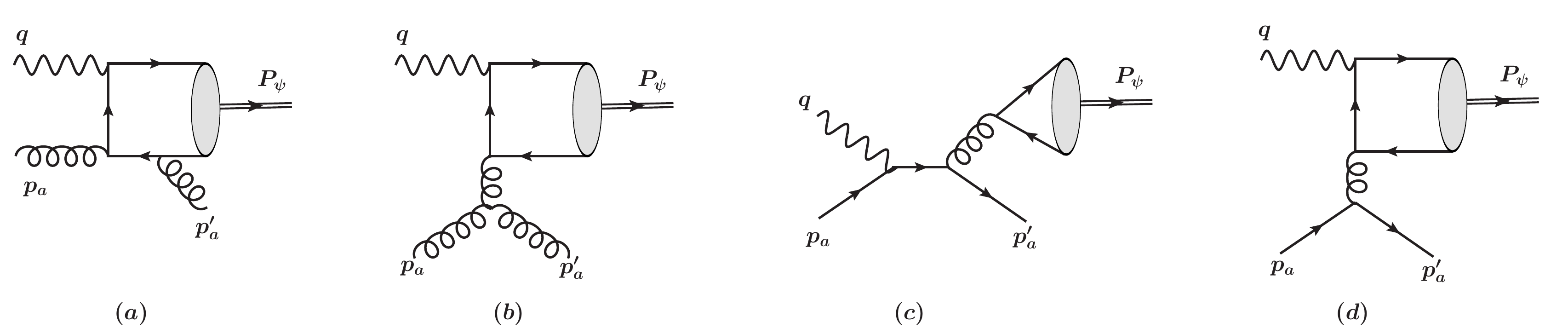} 
\par\end{centering}
\caption{{Representative diagrams of the partonic subprocesses contributing to $J/\psi$ production in SIDIS at the order $\alpha\alpha_s^2$, $\gamma^*(q) + a(p_a) \to {J/\psi}(P_\psi) + a(p_a^\prime)$ with $a=g,q, \bar q$. The six diagrams of type (a)  are the only ones corresponding to the CS production mechanism. Moreover, there are two diagrams for each type (b), (c), (d). }}
\label{fig:fd}
\end{figure}

The helicity structure functions ${\cal W}^{\cal P}_{\lambda\lambda^\prime}$ can be calculated within NRQCD and collinear factorization in the kinematic region $q_\sT \gg \Lambda_{\text{QCD}}$. One first needs  to evaluate the partonic structure functions  ${w}^{{\cal P} (a)}_{\lambda\lambda^\prime}$ for  each of the underlying hard scattering subprocesses, which at  the order $\alpha \alpha_s^2$ are 
\begin{align}
\gamma^* (q) + a(p_a) \to c \overline c[n](P_\psi) + a(p_a^\prime)\,, 
\label{eq:parton-proc}
\end{align}
where  $a$  is either a gluon, a quark or an antiquark with helicity $\lambda_a$, and the charm-anticharm quark pair is produced in the intermediate Fock state $n = \, ^{2S+1\!}L_J^{[c]}$, with $S$, $L$, $J$ being the spin, orbital and total angular momenta of the $c\bar c$ pair, respectively, while $c = 1, 8$ specifies its color configuration.  In addition to the leading color-singlet production channel with $n= \,^3\!S_1^{[1]}$, we also include the subleading color-octet channels that are relatively suppressed by at most a factor of $v^4$, namely $^1\!S_0^{[8]}$,  $^3\!S_1^{[8]}$, $^3\!P_J^{[8]}$, with $J=0,1,2$. The corresponding Feynman diagrams are shown in Fig.~\ref{fig:fd}. The polarized hadronic structure functions are then obtained as convolutions of the partonic ones with suitable parton distribution functions $f_1^a$. 
If we denote by  ${\cal M}_{\mu\alpha} (\gamma^* a\to J/\psi \, X)$ the amplitude for the processes $\gamma^* a\to J/\psi \, X$,  where we assume  that the $J/\psi$ meson is produced in a definite helicity state $\lambda$,  the partonic structure functions are given by
\begin{align}
w^{{\cal P} (a)}_{\lambda\lambda^\prime} = \sum_{\lambda_a}\,\frac{1}{2} \, \varepsilon^{\mu\nu}_{\cal P}\,{\cal M}_{\mu\alpha}(\gamma^* a \to J/\psi\,X)\,{\cal M}^{*}_{\nu\beta}(\gamma^* a \to J/\psi\,X)\, \epsilon^\alpha_{\lambda}(P_\psi) \,  \epsilon^{* \beta}_{\lambda^\prime}(P_\psi)  \,,
\label{eq:hel-part}
\end{align}
with ${\cal{P}} = \perp,\myparallel$; the tensors $\varepsilon^{\mu\nu}_{\cal P}$ being defined as $\varepsilon^{\mu\nu}_{\perp} = -g^{\mu\nu}_\perp$ and $\varepsilon^{\mu\nu}_\myparallel = \epsilon^{\mu}_{\gamma \myparallel}\, \epsilon^{\nu}_{\gamma \myparallel}$, see also Eqs.~\eqref{eq:gperp}-\eqref{eq:epsL-gamma}. The  $J/\psi$ polarization vectors $\epsilon_\lambda^\alpha (P_\psi)$ are presented in Appendix~\ref{sec:frames} in terms of the four-momenta  $q$, $P$, $P_\psi$ in the four different frames considered. 

According to NRQCD, the $J/\psi$ polarization state $\lambda$ can be reached from the $c\bar c$ pairs produced in various orbital and spin angular momentum states in the subprocesses of Eq.~\eqref{eq:parton-proc}~\cite{Beneke:1998re}.  Because of charge  and parity conjugation, there are no interference effects between  intermediate states with different orbital angular momentum $L$  and spin $S$. However, NRQCD does not forbid interferences of different $^3P_J$ states,  for which $S=L=1$~\cite{Leibovich:1996pa,Beneke:1995yb,Beneke:1996tk}. The following decomposition of $w^{{\cal P} (a)}_{\lambda\lambda^\prime} $ is therefore valid, 
 \begin{align}
w^{{\cal P} (a)}_{\lambda\lambda^\prime} =w^{{\cal P} (a)}_{\lambda\lambda^\prime}\left [\,^3\!S_1^{[1]} \right ] + w^{{\cal P} (a)}_{\lambda\lambda^\prime}\left [\,^3\!S_1^{[8]} \right ] \,+\,  w^{{\cal P} (a)}_{\lambda\lambda^\prime}\left [\,^1\!S_0^{[8]} \right ]
\,+\,  w^{{\cal P} (a)}_{\lambda\lambda^\prime} \left [ \left \{ L = 1, S=1 \right \}^{[8]} \right ] \,,
\label{eq:w-dec}
\end{align}
which expresses the coherent sum of the partonic structure functions for the  four intermediate $c\bar c$ states.
Since the $^1S_0^{[8]}$ state is rotational invariant in the quarkonium rest frame, it leads to a random orientation of the $J/\psi$ spin. In other words, each $J/\psi $ helicity state $\lambda$ amounts to one third of the unpolarized cross section. Furthermore, for the $^3\!S_1^{[1,8]}$ states, NRQCD spin symmetry implies that the perturbatively calculable third component of the spin of the $c \bar c$ pair, namely the quantum number $S_z$, is not modified in the subsequent nonperturbative evolution into the observed $J/\psi$ meson. Hence, $S_z$ is equal to the helicity of the $J/\psi$, {\it i.e.}\ 
$S_z =\lambda$. The  only nontrivial term of the sum in Eq.~\eqref{eq:w-dec} is the last one. This can be calculated explicitly by projecting the hard scattering amplitude onto states of definite $S_z=\lambda$ and $L_z$, squaring the amplitude and summing over $L_z$~\cite{Beneke:1998re,Leibovich:1996pa,Beneke:1995yb}. In formulae, 
\begin{align}
w^{{\cal P} (a)}_{\lambda\lambda^\prime} \left [ \left \{ L = 1, S=1 \right \}^{[8]}  \right ]& \propto   \sum_{L_z}  \sum_{\lambda_a}\,\frac{1}{2} \, \varepsilon^{\mu\nu}_{\cal P}\,{\cal M}_{\mu\alpha} \left  (\gamma^* a \to c\bar c[(1, L_z; 1,\lambda)]\, a \right ) \, 
{\cal M}^{*}_{\nu\beta}\left  (\gamma^* \to c\bar c[(1, L_z; 1,\lambda^\prime)\,a] \right ) \epsilon^\alpha_{\lambda} \,  \epsilon^{* \beta}_{\lambda^\prime} \nonumber \\
& \neq \sum_{J=0,1,2} w^{{\cal P} (a)}_{\lambda\lambda^\prime}[\,^3P_J]\,,
\label{eq:K-3P0}
\end{align}
where we have denoted the quantum numbers of the $c\bar c $ pair by $(L,L_z;S,S_z)$. At the order in $v$ we are considering and by further exploiting the symmetries of NRQCD, it can be shown that the partonic structure functions can be written in a factorized form, in which the nonperturbative contributions, encoded in the same four LDMEs   
$\langle 0 \vert {\cal O} (n)\vert 0 \rangle$ that describe unpolarized $J/\psi$ production~\cite{Boer:2020bbd,Kniehl:2001tk,Sun:2017nly,Sun:2017wxk,Zhang:2019ecf}, are multiplied by perturbative  short-distant coefficients $K^{{\cal P} (a)}_{\alpha\beta} [n]$. These can be calculated from the hard scattering partonic subprocesses $\gamma^* a \to c \bar c[n]\, a$. Hence we can write
\begin{align}
w^{{\cal P} (a)}_{\lambda\lambda^\prime} = \sum_n w^{{\cal P} (a)}_{\lambda\lambda^\prime}[n] \equiv  \sum_n K^{{\cal P} (a)}_{\alpha\beta} [n] \, \epsilon^\alpha_{\lambda}(P_\psi) \, \epsilon^{* \beta}_{\lambda^\prime}(P_\psi) \,\langle 0 \vert {\cal O} (n)\vert 0 \rangle \,,
\label{eq:w-fact}
\end{align}
where the sums run over the Fock states $n=\,^3\!S_1^{[1]}$, $^1\!S_0^{[8]}$,  $^3\!S_1^{[8]}$, $^3\!P_0^{[8]}$. 

In the actual calculation we neglect the proton mass and any smearing effects both in the initial and in the final states, hence partons can be taken to be collinear to the parent proton, $p_a^{\mu} = \xi P^\mu$. The  $w^{{\cal P} (a)}_{\lambda\lambda^\prime}$ can therefore be expressed as functions of the scalar products of the four-vectors $q$, $p_a$, $P_\psi$, or, equivalently, of the usual partonic Mandelstam variables
\begin{align}
\hat s = (q+ p_a)^2\,,\qquad \hat t = (q-P_\psi)^2\,,\qquad \hat u = (p_a-P_\psi)^2\,. 
\label{eq:mandelstam}
\end{align}
It is then straightforward to obtain the structure functions $w^{{\cal P} (a)}_{\Lambda}$, with $\Lambda = T,L, \Delta, \Delta\Delta $, from the $w^{{\cal P} (a)}_{\lambda\lambda^\prime}$ in analogy to Eq.~\eqref{eq:W-P-L}. 
Their expressions in the GJ frame are collected in Appendix~\ref{sec:helicity}, while the corresponding formulae in the other three reference frames considered (TF, CS, HX) can be obtained by applying the transformations listed  in Appendix~\ref{sec:frames}. We note that in the photoproduction limit ($Q\to 0$) our results agree with the ones in Ref.~\cite{Beneke:1998re}. Furthermore, we are able to reproduce the SIDIS unpolarized cross section, related to the combination $2W_T + W_L$,  presented in Ref.~\cite{Kniehl:2001tk}.  The structure functions $W_T$ and $W_L$ for $J/\psi$ production in SIDIS have been calculated also in Ref.~\cite{Yuan:2000cn}, however we are in disagreement with the relative normalizations of most of the contributions to the structure functions for longitudinally polarized virtual photons, $w^{{\myparallel} (g)}_{\Lambda}[n]$, presented in Ref.~\cite{Yuan:2000cn}.
Since  $w^{{\myparallel} (g)}_{\Lambda}[n] \to 0$ as $Q\to 0$,  this might explain why the authors of Ref.~\cite{Yuan:2000cn}  recover the photoproduction results in Ref.~\cite{Beneke:1998re}, but not the unpolarized SIDIS cross section in Ref.~\cite{Kniehl:2001tk}. 
Conversely, to the best of our knowledge, the SIDIS helicity flip structure functions $W_{\Delta}$ and $W_{\Delta\Delta}$ have been explicitly computed for the first time in the present work.

In order to write down the expressions for the hadronic structure functions, it is convenient to  introduce the scaling variables
\begin{equation}
\hat{x} = \frac{Q^2}{2p_a\cdot q}\,,\qquad \hat z = \frac{p_a\cdot P_\psi}{p_a\cdot q}\,.
\end{equation}
By comparison with the hadronic variables in  Eq.~\eqref{eq:SIDIS-var},  we obtain the relations $\hat z = z$ and $\hat x = {\xB}/{\xi}$. The latter implies $\hat x p_a = \xB P$, with $\hat x \ge \xB$. 
These variables allow us to perform, in the $J/\psi$ rest frame, the following Sudakov decomposition of the four-momenta of the particles involved in the reaction, 
\begin{align}
q^\mu  & = \frac{M_\psi}{\sqrt{2} \hat z} \, n_+^\mu  - \left (1-\frac{\bm q_\sT^2}{Q^2}   \right )\frac{\hat z \,Q^2}{\sqrt{2} M_\psi}\, n^\mu_-  + q_\sT^\mu\,, \nonumber \\
p_a^\mu & = \frac{\hat z\, Q^2}{\sqrt{2} \, \hat x\, M_\psi}\, n_-^\mu\,, \nonumber \\
P^\mu_\psi & = \frac{M_\psi}{\sqrt{2}}\, n^\mu_+ + \frac{M_\psi}{\sqrt{2}}\, n^\mu_-\,,
\end{align}
where $n_+$ and $n_-$ are two light-like vectors such that $n_+\cdot n_-=1$ and $\bm q_\sT^2= -q_\sT^2$.  Hence the Mandelstam variables in Eq.~\eqref{eq:mandelstam} can be written as
\begin{align}
\hat s =  \frac{1-\hat x}{\hat x}\, Q^2\,,\qquad \hat t =-(1- \hat z)
 \left ( Q^2 + \frac{M^2_\psi}{\hat z} \right ) - \hat z \bm q_\sT^2\, ,
 \qquad  \hat u =   -\frac{\hat z}{\hat x}\, Q^2 + M_\psi^2 \,,
\end{align}
and the hadronic structure functions are given by the convolutions
\begin{align}
(2\pi)^3 {\cal W}^{\cal P}_{\lambda \lambda^\prime}& = \sum_a \int_{\xB}^{\hat x_{\rm max}}
 \frac{\d \hat x}{\hat x}\int_z^1  \d \hat z\,   f_1^a \left (\frac{\xB}{ \hat x} \,,\mu^2\right ) \,w^{{\cal P} (a)}_{\lambda\lambda^\prime} \,  \delta \left (\hat s + \hat t + \hat u - M_\psi^2 + Q^2 \right )
 \delta (\hat z-z) \nonumber  \\
& =   \sum_{a} \sum_n \int_{\xB}^{\hat x_{\rm max}}
 \frac{\d \hat x}{\hat x}\int_z^1  \frac{\d \hat z}{\hat z}\,  \frac{1}{Q^2}  \, f_1^a \left (\frac{\xB}{ \hat x} \,,\mu^2\right ) \, 
 w^{{\cal P} (a)}_{\lambda \lambda^\prime}[n]
 \nonumber \\
 &\qquad \qquad  \qquad \times  \delta \left ( \frac{\bm q_\sT^2}{Q^2}  + \frac{1-\hat z}{\hat z^2 }\, \frac{M_\psi^2}{Q^2} -\frac{(1-\hat x)(1- \hat z) }{\hat x \hat z}  \right )\delta(\hat z-z)\,,
 \label{eq:W-had-coll-1}
\end{align}
where the final result has been obtained  by substituting the expression for  $w^{{\cal P} (a)}_{\lambda\lambda^\prime}$ given  in Eq.~\eqref{eq:w-fact}, while
\begin{align}
\hat x_{\rm max} = \frac{Q^2}{Q^2 + M_\psi^2}\,.
\end{align}
Moreover, in Eq.~\eqref{eq:W-had-coll-1} $\mu^2$ is the hard factorization scale, on which also the partonic structure functions  $w^{{\cal P} (a)}_{\lambda\lambda^\prime}[n]$ depend, even if not explicitly shown. 

The helicity structure functions discussed so far are expected to describe 
the azimuthal decay distributions of $J/\psi$ mesons for large values of the transverse momentum, namely for $\bm q_\sT^2 \gg \Lambda^2_\text{QCD}$. Their behavior in the small-$q_\sT$ region, $\Lambda^2_\text{QCD} \ll \bm q_\sT^2\ll Q^2$, can be obtained along the lines of Ref.~\cite{Boer:2020bbd}, by replacing the Dirac delta in Eq.~\eqref{eq:W-had-coll-1} with its expansion in the small-$q_\sT$ limit,
\begin{align}
\delta \left (\frac{\bm q_\sT^2}{Q^2} + \frac{1-\hat z}{\hat z^2} \, \frac{M_\psi^2}{Q^2}- \frac{(1-\hat x)(1- \hat z)}{\hat x \hat z} \right ) &  \approx \hat x_{\rm max} \left \{\frac{\hat x^\prime }{(1- \hat  x^\prime)_+} \, \delta (1-\hat z) \, +
\,   \frac{Q^2 + M^2_\psi}{Q^2 + M_\psi^2/\hat z}   \,  \frac{\hat z}{(1-\hat z)_+} \ \delta \left (1 -\hat x^\prime \right )  \right .\nonumber \\
&\qquad  \left .+ \,  \delta (1- \hat x^\prime) \delta (1-\hat z)  \ln \bigg(\frac{Q^2+M_\psi^2}{\bm q_\sT^2}   \bigg) \right \}\,,
\end{align}
where $\hat x^\prime = {\hat x}/{\hat x_{\rm max}}$\,. Up to corrections of the order of  ${\cal O} ( \Lambda_{\rm QCD} / \vert \bm q_\sT\vert)$ and ${\cal O} (\vert \bm q_\sT\vert/Q)$, we find the following leading power behavior of the structure functions for transversely polarized photons
\begin{eqnarray}
{\cal W}_\Lambda^{\perp} & = &  \widetilde{w}_\Lambda^{\perp \,(g)}\, \frac{\alpha_s}{2 \pi^2 \bm q_\sT^2} \left[ L \left( {Q^2 + M_\psi^2 \over {\bm q}_\sT^2} \right) f_1^g(x,\mu^2) + \left( P_{gg} \otimes f_1^g + P_{gi} \otimes f_1^i \right)(x,\mu^2) \right]  \qquad \text{with}\quad \Lambda = T, L, \nonumber \\
{\cal W}_{\Delta\Delta}^{\perp} &= & \widetilde{w}_{\Delta\Delta}^{\perp\,(g)}  \,\frac{\alpha_s}{\pi^2   \bm q_\sT^2} \left( \delta P_{gg} \otimes f_1^g + \delta P_{gi} \otimes f_1^i \right) (x,\mu^2) \,,
\label{eq:wPT}
\end{eqnarray}
where a sum over $i = q, \bar{q}$ is understood. The single-helicity flip structure function  ${\cal W}_{\Delta}^\perp$ is suppressed by a factor $\vert \bm q_\sT\vert/Q$ relatively to the ones in Eq.~\eqref{eq:wPT} and therefore we do not consider it in the following. Similarly, for longitudinally 
polarized photons, 
\begin{align}
{\cal W}_L^{\myparallel}  =  \widetilde{w}_L^{\myparallel\, (g)} \,\frac{\alpha_s}{2 \pi^2  \bm q_\sT^2} \left[ L \left( {Q^2 + M_\psi^2 \over {\bm q}_\sT^2} \right) f_1^g(x,\mu^2) + \left( P_{gg} \otimes f_1^g + P_{gi} \otimes f_1^i \right)(x,\mu^2) \right]\,,
\label{eq:WL-smallqT}
\end{align}
while all the other structure functions are suppressed. We note that the above results are independent of the chosen reference frame. The quantities  $\widetilde{w}_\Lambda^{{\cal P}\, (g)}$  in Eqs.~\eqref{eq:wPT} and~\eqref{eq:WL-smallqT} are  the gluon helicity structure functions at LO in $\alpha_s$, {\it i.e.}\ they refer to the $2\to 1$ subprocess $\gamma^* g \to J/\psi$ in Fig.~\ref{fig:fd-lo} and  read 
\begin{align}
\widetilde {w}_T^{\perp\,(g)} & =  2 (4 \pi)^2  {\alpha \alpha_s e_c^2 \over   M_\psi  (M_\psi^2+Q^2)}   \left[ {1 \over 3} \langle 0|{\cal O}_8 (^1 S_0)|0\rangle + {4 \over M_\psi^2} {3 M_\psi^4 + Q^4 \over (M_\psi^2 + Q^2)^2} \langle 0|{\cal O}_8 (^3 P_0)|0\rangle \right]\delta (1-\hat z)\,, \nonumber \\
\widetilde  w_L^{\perp\,(g)} & =  2 (4 \pi)^2 {\alpha \alpha_s e_c^2 \over   M_\psi   (M_\psi^2+Q^2)} \left[ {1 \over 3} \langle 0|{\cal O}_8 (^1 S_0)|0\rangle 
+ {4 \over M_\psi^2} \langle 0|{\cal O}_8 (^3 P_0)|0\rangle \right] \delta (1-\hat z)\,, \nonumber \\
\widetilde{w}_{\Delta\Delta}^{\perp\, (g)} & =  - 16 (4 \pi)^2 {\alpha \alpha_s e_c^2 \over   M_\psi (M_\psi^2 + Q^2)^2}  \langle 0|{\cal O}_8 (^3 P_0)|0\rangle \, \delta (1-\hat z)\,,\nonumber \\
 \widetilde {w}_L^{\myparallel\, (g)} & = {128  (4 \pi)^2} \frac{\alpha\alpha_s e_c^2\, Q^2}{M_\psi (M_\psi^2 + Q^2)^3}\, \langle 0|{\cal O}_8 (^3 P_0)|0\rangle \, \delta (1-\hat z)\,,
 \label{eq:wPL}
\end{align}
where, as before, the superscripts (subscripts) refer to the photon ($J/\psi$) polarization states  and $e_c$ is the electric charge of the charm quark in units of the proton charge. Furthermore, in Eqs.~\eqref{eq:wPT} and~\eqref{eq:WL-smallqT} we have used the definitions
\begin{align}
x \equiv \frac{\xB}{\hat x_{\rm max}} = \xB\, \left (  1+ \frac{M^2_\psi}{Q^2} \right )  \,,
\end{align}
and
\begin{align}
L \left ( \frac{Q^2+ M^2_\psi}{\bm q_\sT^2} \right ) \equiv 2 C_A\, \ln\left ( \frac{Q^2+ M^2_\psi}{\bm q_\sT^2} \right )- \frac{11C_A-4n_fT_R}{6}\, ,
\label{eq:L-glu}
\end{align}
where $T_R=1/2$,  $n_f$ is the number of active flavors and $C_A= N_c$, with $N_c$ being the number of colors. The symbol $\otimes$ stands for a convolution in the longitudinal momentum fractions of the splitting functions and the parton distributions, 
\begin{align}
(P\otimes f)(x,\mu^2) = \int_x^1\frac{\d\hat x}{\hat x}\, P \left ( \hat x, \mu^2 \right )  f \left ( \frac{x}{\hat x} , \mu^2\right ) \,.
\label{eq:def-conv}
\end{align}
The leading-order (LO) unpolarized splitting functions are explicitly given by
\begin{align}
P_{gg} (\hat x) & = 2 C_A \left [ \frac{\hat x}{(1-\hat x)_+} + \frac{1-\hat x}{\hat x} + \hat x (1-\hat x) \right ] + \delta (1-\hat x) \, \frac{11C_A-4n_f T_R}{6}\,,\nonumber \\
P_{gq} (\hat x)& = P_{g\bar q}  (\hat x) =C_F \, \frac{1+ (1-\hat x)^2}{\hat x}\,,
\label{eq:spl-funct}
\end{align}
with $C_F = (N_c^2-1)/2 N_c$. As well known, the plus-prescription on the singular parts of the splitting function $P_{gg}$ is defined so that the integral of a sufficiently smooth distribution $G$ is given by 
\begin{align}
\int_z^1\d y \,  \frac{G(y)}{(1-y)_+} = \int_z^1\d y\,  \frac{G(y)-G(1)}{1-y} - G(1) \ln \left ( \frac{1}{1-z} \right )\,.
\label{eq:plus}
\end{align}
Moreover, the splitting functions of an unpolarized parton into a linearly polarized gluon read~\cite{Sun:2011iw,Catani:2010pd} 
\begin{align}
\delta P_{gg}(\hat x) & = C_A\, \frac{1-\hat x}{\hat x}\,, \nonumber \\
\delta P_{gq}(\hat x) & = \delta P_{g\bar q}(\hat x)= C_F\, \frac{1-\hat x}{\hat x}\,.
\label{eq:lin-spl-funct}
\end{align} 

Finally, we point out that the partonic subprocesses which contribute to the structure functions at  small $q_\sT$,  see Eqs.~\eqref{eq:wPT}-\eqref{eq:WL-smallqT}, are only the ones in the intermediate Fock states  $n =~ ^1\!S_0^{[8]}$, $^3\!P_J^{[8]}$, corresponding to $t$-channel Feynman diagrams of the (b) and (d) type in Fig.~\ref{fig:fd}.  The other partonic channels, {\it i.e.}\ the gluon induced $^3\!S_1^{[1,8]}$ subprocesses in Fig.~\ref{fig:fd}~(a) and the quark-induced subprocesses in  Fig.~\ref{fig:fd}~(c), are suppressed and vanish in the limit  $\bm q_\sT^2 \to 0$. This means that they are not relevant in our analysis of the matching of the collinear and TMD results. Hence they will not be considered in the following discussion. 

\section{TMD factorization and matching}
\label{sec:TMD}

\begin{figure}[t]
\begin{centering}
\includegraphics[trim={0cm 0cm 0cm 0cm},clip,scale=0.7]{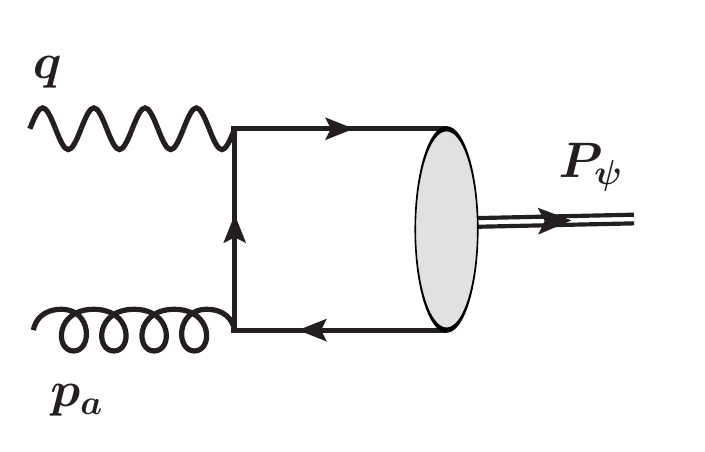}
\par\end{centering}
\caption{Feynman diagram for  the process $\gamma^* (q) \,+ \, g(p_a)\to J/\psi (P_{\psi})$ contributing  to $J/\psi$ production in SIDIS at the order $\alpha\alpha_s$. The crossed diagram, in which the directions of the arrows are reversed, is not shown. 
}
\label{fig:fd-lo}
\end{figure}

Within the TMD framework at the order $\alpha\alpha_s$ in the  strong coupling constant  and 
$v^4$ in the NRQCD velocity parameter, the  underlying partonic subprocesses are given by
\begin{align}
\gamma^* (q) + g(p_a) \to c \overline c[n](P_\psi)\,, 
\end{align}
where the $c\bar c$ pairs are produced perturbatively in one of the color-octet bound states $n =\,  ^1S_0^{[8]}, \,  ^3P_J^{[8]}$ with $J=0,1,2$, see also Fig.~\ref{fig:fd-lo}. 
The leading-twist helicity structure functions for incoming transversely polarized photons read
\begin{align}
{ {\cal W}}_T^{\perp} & =  {2 (4 \pi)^2 {\alpha \alpha_s e_c^2 \over  M_\psi (M_\psi^2 + Q^2)}} \left \{  {1 \over 3}\,  {\cal C} \big [f_1^g \, \Delta_{T}^{[^1S_0]} \big ] (x, \bm q_\sT^2; \mu^2) \, + \, {4 \over M_\psi^2} {3 M_\psi^4 + Q^4 \over (M_\psi^2 + Q^2)^2} \,  {\cal C} \big [f_1^g \, \Delta_{T}^{[^3P_0]} \big ] (x, \bm q_\sT^2; \mu^2)  \right \} \,,\nonumber \\
{ {\cal W}}_L^{\perp} & =  {2 (4 \pi)^2 \frac{\alpha \alpha_s e_c^2}{ M_\psi (M_\psi^2 + Q^2)} }\,\left \{  {1 \over 3}  \,  {\cal C} \big [f_1^g \, \Delta_{L}^{[^1S_0]} \big ] (x, \bm q_\sT^2; \mu^2) \, + {4 \over M_\psi^2} \, {\cal C} \big [f_1^g \, \Delta_{L}^{[^3 P_0]} \big ] (x, \bm q_\sT^2; \mu^2)\rangle  \right \}   \,, \nonumber  \\
{ {\cal W}}_{\Delta\Delta}^{\perp} & =  - { 16 (4 \pi)^2 \frac{\alpha \alpha_s e_c^2}{ M_\psi (M_\psi^2 +\ Q^2)^2}}\, {\cal C} \big [w \, h_1^{\perp g} \, \Delta_{\Delta\Delta}^{[^3P_0] } \big ] (x, \bm q_\sT^2; \mu^2) \,,
\label{eq:WT-small-qT}
\end{align}
while for incoming longitudinally polarized photons we find
\begin{align}
{ {\cal W}}_L^{\parallel} & =  {128 (4 \pi)^2} {\alpha \alpha_s e_c^2\, Q^2\over M_\psi (M_\psi^2 + Q^2)^3}\,  {\cal C} \big [f_1^g \, \Delta_{L}^{[^3P_0]} \big ] (x, \bm q_\sT^2; \mu^2) \,,
\label{eq:WL-small-qT}
\end{align}
where we have introduced the transverse momentum convolutions
\begin{align}
{\cal C} \big [f_1^g \, \Delta_{\Lambda}^{[n]} \big ] (x, \bm q_\sT^2; \mu^2) & = \int \d^2 \bm p_\sT  \int \d^2 \bm k_\sT \,  \delta^2 (\bm q_\sT -\bm p_\sT - \bm k_\sT  )\,  f_1^g (x, \bm p_\sT^2;\mu^2)  \, \Delta_{\Lambda}^{[n]} (\bm k_\sT^2, \mu^2)\,, \nonumber \\
{\cal C} \big [ w \, h_1^{\perp g} \, \Delta_{\Lambda}^{[n]} \big ] (x, \bm q_\sT^2;\mu^2) & = \int \d^2 \bm p_\sT  \int \d^2 \bm k_\sT \,  \delta^2 (\bm q_\sT -\bm p_\sT - \bm k_\sT  )\, w(\bm p_\sT, \bm k_\sT)\,  h_1^{\perp g} (x,  \bm p_\sT^2;\mu^2)  \, \Delta_{\Lambda}^{[n]} ( \bm k_\sT^2,\mu^2)\, .
\label{eq:conv}
\end{align}
In the above equations, $f_1^g$ and $h_1^{\perp\, g}$ are, respectively,  the unpolarized and linearly polarized gluon TMDs inside an unpolarized proton~\cite{Mulders:2000sh,Meissner:2007rx,
Boer:2016xqr,Echevarria:2015uaa,Gutierrez-Reyes:2019rug,Luo:2019bmw}, while $ \Delta_{\Lambda}^{[n]}$ are the shape functions~\cite{Echevarria:2019ynx,Fleming:2019pzj} encoding the smearing  effects in the $c \bar c [n] \to J/\psi$ transition. As such, the $\Delta_\Lambda^{[n]}$ could in principle depend on the quantum numbers of the intermediate $c\bar c$ pair, as well as on the polarization $\Lambda$ of the final $J/\psi$ meson. Moreover, $w(\bm p_\sT, \bm k_\sT)$ is the transverse momentum dependent weight function~\cite{Boer:2020bbd}
\begin{align}
 w(\bm p_\sT, \bm k_\sT) = \frac{1}{ M_p^2 \, \bm q_\sT^2}\, \left [ (\bm p_\sT \cdot \bm q_\sT)^2 -\frac{1}{2}\, \bm p_\sT^2\, \bm q_\sT^2 \right ] \,.
\end{align}

In absence of smearing effects, the final $J/\psi$ meson would be  collinear to the $c\bar c$ pair originally produced in the hard scattering process and the shape functions would be given by $\Delta_{\Lambda}^{[n]} (\bm k_\sT^2;\mu^2) =  \langle 0 \vert {\cal O} (n) \, \vert 0 \rangle \, \delta^2(\bm k_\sT)$.  Hence  the  convolutions in Eq.~\eqref{eq:conv} reduce to the products of a LDME $ \langle 0 \vert {\cal O} (n) \, \vert 0 \rangle$ with a gluon TMD, namely  
\begin{align}
{\cal C} \big [f_1^g \, \Delta_{\Lambda}^{[n]} \big ] (x, \bm q_\sT^2; \mu^2)  & = \langle 0 \vert {\cal O} (n) \, \vert 0 \rangle \,  f_1^g  (x, \bm q_\sT^2; \mu^2)\,, \nonumber \\
 {\cal C} \big [ w \, h_1^{\perp g} \, \Delta_{\Lambda}^{[n]} \big ] (x, \bm q_\sT^2;\mu^2) & =  \langle 0 \vert {\cal O} (n) \, \vert 0 \rangle\,  \frac{\bm q_\sT^2}{2 M_p^2 }\, h_1^{\perp g} (x,  \bm q_\sT^2;\mu^2) \,,\qquad \text{with}~~ \bm q_\sT = \bm p_\sT\,.
 \label{eq:conv-nosm}
\end{align}

TMDs can be calculated perturbatively in the limit $ \vert \bm q_\sT \vert \gg\Lambda_{\rm QCD}$, where their soft parts can be safely neglected. In particular, the unpolarized gluon TMD distribution at the scale $\mu^2$ and order $\alpha_s$ can be expressed in terms of collinear parton distributions as follows~\cite{Boer:2020bbd},
\begin{align}
f_1^g(x, \bm q_\sT^2; \mu^2) 
& = \frac{\alpha_s}{2\pi^2 \bm q_\sT^2}\, \,\left [ \left (C_A  \ln \frac{\mu^2}{ \bm q_\sT^2}  -  \frac{11 C_A- 2 n_f }{6}\,    \right )    f_1^g(x, \mu^2) \,+\,(P_{gg} \otimes f_1^g  \,+\,  P_{gi}\otimes f_1^i) (x, \mu^2) \right ]  \,,
\label{eq:TMD-f1}
\end{align}
where $ i = q, \bar q$, the LO splitting functions $P_{g g}$ and $P_{gi}$ are given in Eq.~\eqref{eq:spl-funct} and the convolution $\otimes$ is defined in Eq.~\eqref{eq:def-conv}. Similarly,  the perturbative tail  of the linearly polarized gluon distribution reads~\cite{Sun:2011iw}
\begin{align}
 \frac{ \bm q_\sT^2}{2 M_p^2}\,h_1^{\perp g } (x,  \bm q_\sT^2; \mu^2) 
 & =  \frac{\alpha_s}{\pi^2} \,  \frac{1}{\bm q_\sT^2}\,(\delta P_{gg} \otimes f_1^g  \,+\,  \delta P_{gi}\otimes f_1^i) (x, \mu^2)\,,
 \label{eq:h1p}
\end{align}
with the LO polarized splitting functions $\delta P_{gg}$ and $\delta P_{gi}$ given in Eq.~\eqref{eq:lin-spl-funct}.  

By substituting Eqs.~\eqref{eq:conv-nosm} together with the expressions for $f_1^g(x, \bm q_\sT^2; \mu^2)$ and $h_1^{\perp g } (x,  \bm q_\sT^2; \mu^2)$  in Eqs.~\eqref{eq:TMD-f1}-\eqref{eq:h1p}, evaluated at the scale $\mu^2 = Q^2+M_\psi^2$, into Eqs.~\eqref{eq:WT-small-qT}-\eqref{eq:WL-small-qT}, we find that, if  smearing is neglected in the final state, only the double-helicity flip structure function ${\cal W}^{\perp}_{\Delta\Delta}$  exactly matches the corresponding collinear one in the small-$q_\sT$ limit given in Eqs.~\eqref{eq:wPT}, \eqref{eq:wPL}.  This is possible without the need of any shape function 
because of the absence of a logarithmic term in ${\cal W}^T_{\Delta\Delta}$ at the perturbative order we are considering.
Conversely, in the intermediate region $\Lambda_{\rm QCD} \ll \vert \bm q_\sT \vert  \ll Q$, 
\begin{align}
\Delta {\cal W}^{\perp}_\Lambda \equiv  {{\cal W}^{\perp}_\Lambda} { \bigg\vert_ {\text{Coll.}}}  \, -  \, {\cal W}^\perp_\Lambda \bigg\vert_ {\text{TMD}} & = \widetilde{w}^{\perp}_{\Lambda}\, \frac{C_A}{2 \pi^2  \bm q_\sT^2}\ln \left ( \frac{Q^2+M^2_\psi}{\bm q_\sT^2} \right )    \qquad \text{with}\quad \Lambda = T, L, \nonumber \\
\Delta {\cal W}^{\myparallel}_L \equiv {\cal W}^{\myparallel}_L \bigg\vert_ {\text{Coll.}}  \, -  \, {\cal W}^{\myparallel}_L \bigg\vert_ {\text{TMD}} & = \widetilde{w}^{\myparallel}_L\, \frac{C_A}{2 \pi^2  \bm q_\sT^2}\,  \ln \left ( \frac{Q^2+M^2_\psi}{\bm q_\sT^2} \right )   \,.
\label{eq:mismatch}
\end{align}

This suggests that smearing effects need to be taken into account in those helicity structure functions which depend on the unpolarized gluon TMD distribution, through the inclusion of a suitable shape function~\cite{Echevarria:2019ynx,Fleming:2019pzj}. Along the lines of the calculation for unpolarized $J/\psi$ production presented in Ref.~\cite{Boer:2020bbd}, to which we refer for details, we are able to find the perturbative tails of the shape functions $\Delta_T^{[n]}$ and $\Delta_L^{[n]}$ at the order $\alpha_s$ by imposing the matching of the TMD and collinear expressions for the  helicity structure functions ${\cal W}^\perp_T$, ${\cal W}^\perp_L$, ${\cal W}^\myparallel_L$ in the intermediate region.  In other words, the shape functions are determined by requiring that their contributions to  ${\cal W}^\perp_T$, ${\cal W}^\perp_L$, ${\cal W}^\myparallel_L$ exactly cancel against  the logarithmic terms on the right-hand side of Eq.~\eqref{eq:mismatch}.  It turns out that $\Delta_T^{[n]} = \Delta_L^{[n]}$, {\it i.e.}\  they are independent of the $J/\psi$ polarization and equal to the shape function  $\Delta ^{[n]}$ calculated for unpolarized $J/\psi$ production in Ref.~\cite{Boer:2020bbd} by applying the same matching procedure,
\begin{align}
\Delta_T^{[n]} ( \bm k_\sT^2,\mu^2) = \Delta_L ^{[n]} ( \bm k_\sT^2,\mu^2) \equiv  \,\Delta ^{[n]} ( \bm k_\sT^2,\mu^2) = \frac{\alpha_s}{2 \pi^2  \bm k_\sT^2}\, C_A \, \langle 0 \vert {\cal O}(n) \vert 0 \rangle  \,  
\ln \frac{\mu^2}{\bm k_\sT^2} \label{tailofshape}\,.
\end{align}
This can be checked directly by plugging into the first line of Eq.~\eqref{eq:conv} the full perturbative expansions of the gluon TMD and the shape function up to the order $\alpha_s$, which are obtained by  adding the order-$\alpha_s^0$ terms to the expressions in Eqs.~\eqref{eq:TMD-f1} and~\eqref{tailofshape}, respectively. We get 
\begin{align}
{\cal C} \big [f_1^g \, \Delta^{[n]} \big ] (x, \bm q_\sT^2; \mu^2) & = \int \d^2 \bm p_\sT  \int \d^2 \bm k_\sT \,  \delta^2 (\bm q_\sT -\bm p_\sT - \bm k_\sT  )\,  f_1^g (x, \bm p_\sT^2;\mu^2)  \, \Delta^{[n]} (\bm k_\sT^2, \mu^2)\nonumber \\
&  = \int \d^2 \bm p_\sT  \int \d^2 \bm k_\sT \,  \delta^2 (\bm q_\sT -\bm p_\sT - \bm k_\sT  )\, \bigg \{  \delta^2(\bm p_\sT)  f_1^g (x, \mu^2)  \nonumber \\
& \qquad  +\,  \left .\frac{\alpha_s}{2\pi^2 \bm p_\sT^2}\, \left [ \left (C_A  \ln \frac{\mu^2}{ \bm p_\sT^2}  -  \frac{11 C_A- 2 n_f }{6}\,    \right )    f_1^g(x, \mu^2) \,+\,(P_{gg} \otimes f_1^g  \,+\,  P_{gi}\otimes f_1^i) (x, \mu^2) \right ] \right \} \nonumber \\
& \qquad \qquad \times \left [  \delta^2(\bm k_\sT) \,+\,  \frac{\alpha_s}{2 \pi^2  \bm k_\sT^2}\, C_A \, \langle 0 \vert {\cal O}(n) \vert 0 \rangle  \,  
\ln \frac{\mu^2}{\bm k_\sT^2} \right ]\,.
\end{align}
Hence, in the limit  $ \bm q_\sT^2 \gg \Lambda_{\text{QCD}}^2,$
\begin{align}
{\cal C} \big [f_1^g \, \Delta^{[n] }\big ] (x, \bm  q_\sT^2)  
& =  \frac{\alpha_s}{2\pi^2  \bm q_\sT^2}\, \langle 0 \vert {\cal O}(n) \vert 0 \rangle\,\left [ \left (2 C_A  \ln \frac{\mu^2}{\bm q_\sT^2}  -  \frac{11 C_A- 2 n_f }{6}\,    \right )    f_1^g(x, \mu^2) \,+\,(P_{gg} \otimes f_1^g  \,+\,  P_{gi}\otimes f_1^i) (x,\mu^2) \, \right ]  \,.
\end{align}
If we substitute this convolution, with the choice $\mu^2=Q^2+M_\psi^2$, into Eqs.~\eqref{eq:WT-small-qT}-\eqref{eq:WL-small-qT}, we recover the correct formulae for ${\cal W}^\perp_T$, ${\cal W}^\perp_L$  and ${\cal W}^\myparallel_L$ in Eqs.~\eqref{eq:wPT}-\eqref{eq:WL-smallqT}. In this way we are able to solve the matching issue of the TMD and collinear results for all the structure functions in the region $\Lambda_{\text{QCD}}^2 \ll  \bm q_\sT^2  \ll Q^2$.  We note that, at the perturbative order we are considering, nothing can be said about the tail of the shape function $\Delta_{\Delta\Delta}^{[n]}$, which  could be in principle different from $\Delta^{[n]}$.  Its determination would require a study  of the matching of the helicity structure function ${\cal W}_{\Delta\Delta}^\perp$ at higher order in $\alpha_s$, because the perturbative expansion of $h_1^{\perp\, g}$ starts at the order $\alpha_s$ and not $\alpha_s^0$ like for $f_1^g$.  However, a full calculation of the cross section for $J/\psi$ production in SIDIS at the order $\alpha^2 \alpha_s^3$ within the NRQCD framework, is still missing. 
 
\section{Summary and conclusions}
\label{sec:conc}

We conclude by summarizing our main results. We have calculated the structure functions ${\cal W}_T, {\cal W}_L, {\cal W}_{\Delta}, {\cal W}_{\Delta\Delta}$, and the corresponding  polarization parameters $\lambda, \mu,\nu$, for $J/\psi$ production in SIDIS.  In particular, the helicity-flip structure functions ${\cal W}_{\Delta}$ and  ${\cal W}_{\Delta\Delta}$, which generate, respectively,  $\cos\phi$ and $\cos2\phi$ azimuthal asymmetries in the angular distribution of the decaying leptons, have been derived here for the first time. In general, the explicit expressions depend on the choice of the coordinate axes. We have presented our results in the Gottfried-Jackson frame and we have shown how to transform them to the Target, Collins-Soper and Helicity frames. 

In the kinematic region where the transverse momentum $q_\sT$ of the virtual photon exchanged in the reaction is large, $q_\sT  \gg \Lambda_\text{QCD}$,  the theoretical frameworks of collinear factorization and NRQCD have been adopted.  
By performing an analysis up to the order $\alpha_s^2$ accuracy, with the inclusion of CO contributions up to  the order $v^4$ with respect to the CS one, we obtain the small-$q_\sT$ behavior of the structure functions ${\cal W}_T$, ${\cal W}_L$ and ${\cal W}_{\Delta\Delta}$ in terms of the LO DGLAP splitting functions.  Furthermore, ${\cal W}_T$ and ${\cal W}_L$ receive also large logarithmic corrections, which are absent in ${\cal W}_{\Delta\Delta}$ and need to be resummed. In contrast, the structure function ${\cal W}_\Delta$ turns out to be suppressed by a factor $q_\sT/Q$ with respect to the other ones. Quite remarkably, these features at small $q_\sT$ do not depend on the choice of the reference frame. 

The large logarithms of the structure functions at small $q_\sT$ can be resummed within the TMD framework. Although a rigorous proof of TMD factorization only exists for light-hadron production in SIDIS, it is reasonable to assume its validity also for the production of heavy quarkonium states, just  changing from the fundamental to the adjoint representation in the gauge-link structure underlying the process~\cite{Bacchetta:2018ivt}.  Hence we propose a factorized expression of the polarized structure functions in terms of transverse momentum dependent parton distributions and shape functions, the latter being  a generalization of the NRQCD LDMEs to TMD approach first introduced in Refs.~\cite{Echevarria:2019ynx,Fleming:2019pzj}.   Our assumption has to fulfill the consistency condition that both descriptions match in the intermediate region 
$\Lambda_\mathrm{QCD} \ll q_\sT \ll \mu$, where $\mu$ is a hard scale typical of the process. At the order $\alpha_s$, we find that a smooth transition from low $q_\sT$ to high $q_\sT$ is possibile if we choose $\mu = \sqrt{Q^2 + M^2_\psi}$.  Moreover, while $W_{\Delta\Delta}$ matches without depending on any shape function because of the absence of logarithmic divergences, from the matching of  $W_L$ and $W_T$ we are able to deduce the specific form of the color-octet shape functions at large transverse momentum.  Very interestingly, the perturbative tails of the shape functions turn out to be independent of the $J/\psi$ polarization, as well as of the quantum numbers of the intermediate color-octet Fock states, except for their overall magnitude given by the NRQCD LDMEs.  These conclusions are in full agreement with the ones obtained in the analysis of unpolarized $J/\psi$ production in SIDIS~\cite{Boer:2020bbd}. They of course hold for any other quarkonium state with the same quantum numbers of the $J/\psi$ meson, such as the $\psi(2 S)$ and $\Upsilon(nS)$ states. 

Finally, our findings show that a combined analysis of the $J/\psi$ polarization parameters and production yields in the large-$q_\sT$  region will be very helpful in testing NRQCD and will likely improve our knowledge of the CO LDMEs.  A similar study in the small-$q_T$ region will provide important information on the shape functions, especially on their nonperturbative part which we cannot predict. Moreover, we suggest a novel experiment for the extraction of the distribution of linearly polarized gluons inside an unpolarized proton, by looking at the double helicity-flip parameter $\nu$, related to  the $\cos 2\phi$ azimuthal decay asymmetry of the $J/\psi$ meson.  The high-luminosity Electron-Ion Collider to be build in the U.S.\ would be the ideal facility  for carrying out this proposal.

\section*{Acknowledgments}
This project has received funding from the European Union's Horizon 2020 research and innovation programme under grant agreement STRONG 2020 - No 824093.  U.D. and C.P.\ also acknowledge financial support by Fondazione di Sardegna under the project {\em Proton tomography at the LHC}, project number F72F20000220007 (University of Cagliari).

\appendix

\section{Reference frames}
\label{sec:frames}

In this appendix, we present  the explicit expressions of the covariant four-vectors  $(T^\mu,X^\mu,Y^\mu,Z^\mu)$, which  form a set of coordinate axes in the four dimensional Minkowski space. These vectors are given as linear combinations of the physical momenta $q^\mu$, $P^\mu$, $P_\psi^\mu$ in four typical reference frames, such that $X^\mu$, $Y^\mu$, $Z^\mu$ become three-vectors in the quarkonium rest frame~\cite{Lam:1978pu,Beneke:1998re}. As already mentioned in Section~\ref{sec:coll}, the normalization conditions are given by  $T^2=1 $ and  $X^2=Y^2=Z^2=-1$, with $T^\mu$ and $Y^\mu$  fixed by Eq.~\eqref{eq:T-Y} in all frames, namely $T^\mu = {P_\psi^\mu}/{M_\psi}$, $Y^\mu = \varepsilon^{\mu\nu\alpha\beta} X_{\nu}Z_\alpha T_\beta$. On the other hand, the definition of $X^\mu$ and $Z^\mu$ is not unique. For their determination, we proceed along the lines of Ref.~\cite{Beneke:1998re}, introducing the following auxiliary four-vectors
\begin{align}
\tilde q^\mu & = q^\mu - \frac{q\cdot P_\psi}{M_\psi^2}\,P_\psi^\mu\,,\nonumber\\
\tilde P^\mu & = P^\mu - \frac{P\cdot P_\psi}{M_\psi^2}\,P_\psi^\mu\,.
\end{align}
Since $\tilde q \cdot T=\tilde P\cdot T =0$, $X^\mu$ and $Z^\mu$ can be easily written as a linear combination of $\tilde q^\mu$ and $\tilde P^\mu$. One starts by writing 
\begin{align}
Z^\mu & = A_z \, \tilde q^\mu+ B_z\, \tilde P^\mu\,,  
\end{align} 
with one of the two coefficients fixed by the condition $Z^2 =-1$. Similarly, 
\begin{align}
X^\mu & = A_x \,\tilde q^\mu + B_x\, \tilde P^\mu\,,  
\end{align}
where $A_x$ and $B_x$ can be determined  by imposing  $X^2 =-1$ and  $X \cdot Z =0$. The remaining sign ambiguity in $A_x$ and $B_x$ is fixed by requiring that the three-vector $\bm Y$ points in the direction of $\bm q \times (- \bm P)$ in the  $J/\psi$ rest frame. 

The four commonly used reference frames described in the following are specified by the choice of $Z^\mu$. The covariant expressions for $X^\mu$ and $Z^\mu$ are given in terms of the  invariants $s = 2  q\cdot p_a$,  $t = M_\psi^2 -Q^2 -2 q\cdot P_\psi$,  $u = M_\psi^2 + Q^2 - 2  p_a \cdot P_\psi $, with $p_a = \xi P$ in the collinear approach.  These invariants are related to the partonic Mandelstam variables in Eq.~\eqref{eq:mandelstam} as follows
\begin{align}
s = \hat s + Q^2\,, \qquad t = \hat{t} \,, \qquad  u = \hat u+ Q^2\,,
\label{eq:invs}
\end{align}
and fulfill the equation
\begin{align}
s+ t + u = Q^2 + M_\psi^2\,.
\end{align}
 
\begin{enumerate}
\item  {\bf Target frame}. The target frame (TF) is defined in such a way that the vector $\bm Z$ points along  the opposite direction of the proton momentum in the $J/\psi$ rest frame, {\it i.e.}\ $\bm Z = - {{\bm P}/|{\bm P}|}$. Hence the four-vectors $Z^\mu$ and  $X^\mu$ are given by 
\begin{align}
	Z^\mu_\text{TF} & = - 2{ M_\psi \over s + t} p_a^\mu + {P_\psi^\mu \over M_\psi} \, , \nonumber \\
	X^\mu_\text{TF} & = - {1 \over (s + t)\sqrt{t (Q^2 t + s u)}} \left\{ (s + t)^2 q^\mu + \left[ s (M_\psi^2 + Q^2) + t (2Q^2 - u) \right] p_a^\mu - s (s + t) P_\psi^\mu \right\}\,.
\end{align}

\item{\bf Collins-Soper frame}. In the Collins-Soper frame (CS),  the direction of $\bm Z$  is chosen as the bisector of the angle between the proton and the virtual photon three-momenta in the $J/\psi$ rest frame, $\bm Z = {{\bf q} / | {\bm q}|} - {{\bm P}/ |{\bm P}|}$. Therefore,
\begin{align}
	Z^\mu_\text{CS} & = {1 \over \sqrt{N_\text{CS}}} \left[ (s + t)\, q^\mu - (s + u - 2Q^2)\, p_a^\mu \right] \, , \nonumber\\
	X^\mu_\text{CS} & = - {1 \over M_\psi} {1 \over \sqrt{t (Q^2 t + su)N_\text{CS}}} \left\{  M_\psi^2 s (s+t)\, q^\mu + M_\psi^2 \left[ s(s+u) +2 Q^2 t \right]\, p_a^\mu - (s + t) \left[ s (M_\psi^2 - t) + Q^2 t \right]\, P_\psi^\mu \right\} ,
\end{align}
with
\begin{equation}
	N_\text{CS} = (s + t) \left[s (M_\psi^2 - t) + Q^2 t\right] \, .
\end{equation}

\item{\bf Gottfried-Jackson frame}.
In the Gottfried-Jackson frame (GJ),  also known as $u$-channel helicity frame, $\bm Z$ is chosen to be along the direction of the three-momentum of the virtual photon in the $J/\psi$ rest frame. Namely, $\bm Z = {{\bm q} / |{\bm q}|}$ and
\begin{align}
	Z^\mu_\text{GJ} & = {1 \over M_\psi \sqrt{N_\text{GJ}}} \left[ 2 M_\psi^2 \, q^\mu - (s + u - 2 Q^2) \, P_\psi^\mu \right] \,, \nonumber  \\
	X^\mu_\text{GJ} & = - {1 \over \sqrt{t (Q^2 t + su) N_\text{GJ}}} \left\{  \left[ s(M_\psi^2 + Q^2) + t(2Q^2 - u) \right]\, q^\mu + \left[ 4 Q^2 t + (s + u)^2 \right]\, p_a^\mu - \left[ 2 Q^2 t + s (s + u) \right]\, P_\psi^\mu \right\} ,
\end{align}
with
\begin{equation}
	N_\text{GJ} = 4 Q^2 t + (s + u)^2\, .
\end{equation}

\item{\bf Helicity frame}.
In the helicity (HX) or recoil frame,  the $Z$-axis is chosen to be  along the direction of the $J/\psi$ three-momentum in the hadronic center of mass frame, that is $\bm Z = - ({\bm P} + {\bm q} )/ | {\bm P} + {\bm q}|$ in the $J/\psi$ rest frame. Moreover,
\begin{align}
	Z^\mu_\text{HX} & = - {1 \over M_\psi \sqrt{N_\text{HX}}} \left\{ 2M_\psi^2 \, (\xi q^\mu + p_a^\mu) - \left[\xi (s + u - 2Q^2) + s + t \right]\, P_\psi^\mu \right\}\,, \nonumber  \\
	X^\mu_\text{HX} & = {1 \over \sqrt{t (Q^2 t + su) N_\text{HX}}} \left\{ \left[ \xi \left(s (M_\psi^2 + Q^2) + t (2 Q^2 - u)\right) - (s + t)^2 \right]\, q^\mu \right.\nonumber\\&
	+ \left[ \xi\left( (s+u)^2 + 4 Q^2  t\right) -s (M_\psi^2 + Q^2) + t (u - 2 Q^2) \right]\, p_a^\mu \nonumber\\&\left.
	+ \left[ s (s + t) - \xi \left(s (s + u) + 2Q^2 t\right) \right]\, P_\psi^\mu \right\} ,
\end{align}
with
\begin{equation}
	N_\text{HX} = \xi^2 \left[ 4 Q^2 t + (s + u)^2 \right] -2\xi \left[ s (M_\psi^2 + Q^2) + t (2Q^2 - u) \right] + (s+t)^2\,.
\end{equation}
\end{enumerate}

The transition from one frame to the other is given by a rotation around the $Y$-axis,
\begin{equation}
\label{eq: pol vec rotation}
\begin{pmatrix}
X \\ Z
\end{pmatrix}_{\! F^\prime}
=
\begin{pmatrix}
\cos \theta & - \sin \theta \\
\sin \theta  &~ ~\cos \theta 
\end{pmatrix}
\begin{pmatrix}
X \\ Z
\end{pmatrix}_{\! F}\,,
\end{equation}
where, from the GJ to the TF frame:
\begin{align}
	\cos \theta_{\text{GJ} \to \text{TF}} & =  \frac{s(M_\psi^2 + Q^2) + t (2 Q^2 - u)}{(s + t) \sqrt{N_\text{GJ}}} \, , \nonumber \\
	\sin\theta_{\text{GJ} \to \text{TF}}   & = \frac{2 M_\psi \sqrt{t (Q^2 t + s u)}}{(s + t) \sqrt{N_{\text{GJ}}}}\,;
\end{align}
from the GJ to the CS frame:
\begin{align}
	\cos \theta_{\text{GJ} \to \text{CS}}   & =  \frac{M_\psi \left[ s (s + u) + 2 Q^2 t \right] }{\sqrt{N_\text{CS} \,N_\text{GJ}}}\, , \nonumber \\
	\sin\theta_{\text{GJ} \to \text{CS}}  & = \frac{(s + u - 2 Q^2) \sqrt{ t (Q^2 t + s u)}}{\sqrt{N_\text{CS} \,N_\text{GJ}}}\, ;
\end{align}
form the GJ to the NX frame:
\begin{align}
	\cos \theta_{\text{GJ} \to \text{HX}} & =  \frac{s(M_\psi^2 + Q^2) + (2Q^2 - u) t - \xi \left[ 4Q^2 t + (s + u)^2 \right] }{\sqrt{N_\text{HX}\, N_\text{GJ}}}\, , \nonumber \\
	\sin\theta_{\text{GJ} \to \text{HX}}  & = \frac{2 M_\psi \sqrt{t (Q^2 t + su)}}{\sqrt{N_\text{HX} \, N_\text{GJ}}}\,.
\end{align}

As a consequence, the hadronic structure functions ${\cal W}_\Lambda$ (as well as the partonic ones) in two different frames $F$ and $F^\prime$ are connected through the linear transformation
\begin{equation}
\begin{pmatrix}
{\cal W}_T \\ {\cal W}_L \\ {\cal W}_\Delta \\ {\cal W}_{\Delta\Delta}
\end{pmatrix}_{\! \! \! F^\prime}
=
\begin{pmatrix}
	1 - {1 \over 2} \sin^2 \theta & {1 \over 2} \sin^2 \theta &  {1 \over 2}\sin 2\theta & {1 \over 2} \sin^2 \theta \\
	\sin^2 \theta & \cos^2 \theta & - \sin 2 \theta& - \sin^2 \theta \\
	- {1 \over 2}\sin 2\theta & {1 \over 2}\sin 2\theta & \cos 2\theta  & {1 \over 2}\sin 2\theta \\
	{1 \over 2} \sin^2 \theta & - {1 \over 2} \sin^2 \theta & - {1 \over 2}\sin 2\theta & 1 - {1 \over 2} \sin^2 \theta
\end{pmatrix}
\begin{pmatrix}
{\cal W}_T \\ {\cal W}_L \\ {\cal W}_\Delta \\ {\cal W}_{\Delta\Delta}
\end{pmatrix}_{\! \! \! F}\,.
 \label{eq: H rotation matrix}
\end{equation}
Similarly, for the polarization parameters we have
\begin{equation}
\begin{pmatrix}
\lambda \\ \mu \\ \nu
\end{pmatrix}_{\!\!\! F^\prime}
=\, 
{1 \over 1 +\rho}
\begin{pmatrix}
	1 - {3 \over 2} \sin^2 \theta & {3 \over 2}\sin 2\theta  & {3 \over 4} \sin^2 \theta \\
	- {1 \over 2}\sin 2\theta & \cos 2 \theta & {1 \over 4}\sin 2\theta \\
	\sin^2 \theta & - \sin 2 \theta & 1 - {1 \over 2} \sin^2 \theta 
\end{pmatrix}
\begin{pmatrix}
\lambda \\ \mu \\ \nu
\end{pmatrix}_{\!\!\! F}
\ , \label{eq: lmn rotation matrix}
\end{equation}
with
\begin{equation}
\rho =  {\sin^2 \theta \over 2} \left( \lambda_F - {\nu_F \over 2} \right) - \sin 2 \theta \, {\mu_F \over 2} \,.
\end{equation}

\section{Partonic helicity structure functions}\label{sec:helicity}

We collect here the analytic expressions for the partonic helicity structure functions in the GJ frame. For each subprocess initiated by a parton $a$, and for each Fock state $n$, they are given in the form
\begin{align}
	w_\Lambda^{{\cal P}\, (a)}[n] = F[n] \, \widehat w_\Lambda^{{\cal P}\, (a)}[n] \,,
\end{align}
where, as already stated, ${\cal P} = \perp,\myparallel$ refers to the virtual photon polarization, while  $\Lambda = T,L,\Delta,\Delta\Delta$ to the $J/\psi$ one.  Moreover, $F[n]$ is a common prefactor independent of the polarizations. 
In order to render the formulae more compact, they are given in terms of the Lorentz invariants $s$, $t$, $u$ in Eq.~\eqref{eq:invs}. Moreover,  for simplicity we will drop the superscript $(a)$ on $\widehat w_\Lambda^{{\cal P}\, (a)}$, as well as the dependences of $F$ and $\widehat w_\Lambda^{\cal P}$ on $n$ when not needed.

\begin{itemize}
\item  $\gamma^* q(\bar q) \,\to \,c \bar c [^1S_0^{[8]}]\, q (\bar q)$:
\begin{align}
	{F}  & =  - { 64\,  e_c^2 \alpha \alpha_s^2   \over 9 s^2 t (s + u)^2 M_\psi }\,  \langle 0 \vert {\cal O}_8(\,^1 S_0) \vert  0 \rangle, \nonumber \\
	\widehat{w}_T^\perp & =  2 Q^2 t \left[ Q^2 t + s (s + u) \right] + s^2 (s^2 + u^2)\,, \nonumber \\
	\widehat{w}_T^\myparallel  & =  8 Q^2 t \left( Q^2 t + s u \right) \, ,\nonumber \\
	\widehat{w}_L^\perp & =\widehat {w}_T^\perp\,,\nonumber \\
	\widehat{w}_L^\myparallel & = \widehat{w}_T^\myparallel\,,\nonumber \\
	\widehat{w}^\perp_{\Delta} & = \widehat {w}^\myparallel_{\Delta}=  \widehat{w}^\perp_{\Delta\Delta} = \widehat{w}^\myparallel_{\Delta\Delta} = 0 \,.
\end{align}

\item  $\gamma^* q(\bar q) \,\to \,c \bar c [^3 S_1^{[8]}]\, q(\bar q)$:
\begin{align}
	F = & -  {16\, e_q^2 \alpha \alpha_s^2  \over 9 s^2 (Q^2 - s)^2 (Q^2 - u)^2 \left[4Q^2t + (s + u)^2\right]  M_\psi^3 } \langle 0 \vert {\cal O}_8(\,^3 S_1) \vert  0 \rangle, \nonumber\\
	\widehat{w}_T^\perp =&  4 Q^{10} t^3 + 4 Q^8 t^2 \left[ s^2 + s (t + u) + t (t - u) \right]  
	+ 2 Q^6 t \left[ s^4 - 2 s^3 (t - 2 u) - s^2 (7 t^2 - 2 t u - u^2) \right.\nonumber\\&\left. - 4 s t^2 (t - u) + t^2 u^2 \right] 
	- 2 Q^4 s \left[ s^4 (t - u) + s^3 u (5 t - 2 u) -  s^2 (2 t^3 - 10 t^2 u - 3 t u^2 + u^3) \right.\nonumber\\&\left. - s t (2 t^3 - 4 t^2 u + u^3) + 2 t^3 u^2 \right] 
	+ Q^2 s^2 \left[ s^4 (t - 2 u) + 2 s^3 (t^2 + t u - 3 u^2) \right.\nonumber\\&\left. + 2 s^2 (t^3 + 3 t^2 u - 2 t u^2 - 3 u^3)  - 2 s u (-2 t^3 - t^2 u + t u^2 + u^3) + t u^2 (2 t^2 - 2 t u - u^2) \right] \nonumber\\&
	+ s^3 u (s^2 + u^2) \left[ s^2 + 2 s (t + u) + 2 t^2 + 2 t u + u^2 \right]
	\,,\nonumber\\
	\widehat{w}_T^\myparallel  = & 8 Q^2 t (Q^2 - s)^2 \left\{ 2 Q^4 t^2 + 2 Q^2 t \left[ 2s^2 + s (3t + u) + t (t - u) \right] + s^2 (s + t + u)^2 + t^2 u^2  \right\} 
	\, ,\nonumber \\
	\widehat{w}_L^\perp = & 4 M_\psi^2 t \left[2 Q^8 t^2 + 2 Q^6 s t (s - 2 t + u) + Q^4 s^2 (s^2 - 2 s t + 2 t^2 - 6 t u + u^2)   \right.\nonumber\\&\left. - 2 Q^2 s^3 u (s - 2 t + u) + 2 s^4 u^2\right]
	\,,\nonumber \\
	\widehat{w}_L^\myparallel = & 32 Q^2 M_\psi^2 t^2 (Q^2 - s)^2 (Q^2 t+s u) 
	\,,\nonumber \\
	\widehat{w}^\perp_{\Delta} = & 2 M_\psi \sqrt{t (Q^2 t + s u)} \left\{ 4 Q^8 t^2+2 Q^6 t \left[ s^2+s (u-3 t)-t u \right] \right.\nonumber\\& -Q^4 s \left[s^3+2 s^2 t+s u (6 t-u)-4 t^2 u\right] 
	+2 Q^2 s^2 t \left[ s (t+u)+u (u-t) \right] \nonumber\\&\left. + s^3 u (s-u) (s+2t+u) \right\}
	\,,\nonumber \\
	\widehat {w}^\myparallel_{\Delta} = & 16 Q^2 M_\psi t \sqrt{t (Q^2 t + s u)} (Q^2 - s)^2 \left[2 Q^2 t + s^2 + s(t+u) - ut \right] 
	\,,\nonumber \\ 
	\widehat{w}^\perp_{\Delta\Delta}  = & 2 t (Q^2 t + s u) \left\{ - 2 Q^8 t + 2 Q^6 \left[ s^2+ 3st + t (t+u) \right] - 4 Q^4 s t (s+t+u) \right.\nonumber\\&\left.  - Q^2 s^2 (s^2+4 s u-2 t^2+u^2) +2 s^3 u (s+t+u) \right\}  
	\,,\nonumber \\ 
	\widehat{w}^\myparallel_{\Delta\Delta} =  & 16 Q^2 M_\psi^2 t^2 (Q^2 t + s u) (Q^2 - s)^2 \,.
\end{align}

\item  $\gamma^* q(\bar q) \,\to \,c \bar c [^3P_J^{[8]}]\, q(\bar q)$:
\begin{align}
	F = & -  {256\, e_c^2 \alpha \alpha_s^2 \over 3 s^2 t (s + u)^4 \left[4Q^2t + (s + u)^2\right] M_\psi^3 } \, \langle 0 \vert {\cal O}_8(\,^3 P_0) \vert  0 \rangle, \nonumber\\
	\widehat{w}_T^\perp = & -16 Q^{10} t^3
	-8 Q^8 t^2 \left[ 3 s^2+2 s u-u (4 t+u) \right] 
	+4 Q^6 t \left[ 2 s^4+s^3 (5 t-2 u) \right.\nonumber\\&\left. +s^2 (10 t^2+7 t u+2 u^2) +s (8 t^3+4 t^2 u+7 t u^2+2 u^3)+t (4 t^3-6 t u^2-3 u^3) \right] \nonumber\\&
	+2 Q^4 \left[ 2 s^6+s^5 (4 u-6 t) -s^4 (t^2+8 t u-4 u^2) +2 s^3 (3 t^3+8 t^2 u-2 t u^2+2 u^3) \right.\nonumber\\&\left. +2 s^2 (2 t^4+9 t^3 u+3 t^2 u^2-4 t u^3+u^4) +2 s t u (4 t^3+t^2 u-4 t u^2-3 u^3)+t^2 u^2 (4 t^2+6 t u+3 u^2) \right] \nonumber\\&
	+2 Q^2 s \left[ -3 s^6+s^5 (t-9 u)  +s^4 (6 t^2+11 t u-12 u^2) +2 s^3 (2 t^3+12 t^2 u+9 t u^2-6 u^3)  \right.\nonumber\\&\left.
	+s^2 u (12 t^3+20 t^2 u+10 t u^2-9 u^3)+s u^2 (4 t^3+8 t^2 u+5 t u^2-3 u^3) +t u^3 (4 t^2+6 t u+3 u^2) \right] \nonumber\\&
	+s^2 (s+u)^2 (s^2+u^2) \left[ 3 s^2+6 s (t+u)+4 t^2+6 t u+3 u^2 \right]
	\,,
	\nonumber
\end{align}
\begin{align}
	\widehat{w}_T^\myparallel  = & 8 Q^2 t \left\{ 
	- 8 Q^8 t^2
	+ 4 Q^6 t \left[s^2-2 s u+u (4 t+u)\right] \right.
	- 2 Q^4 \left[ s^3 (3 t-2 u)+s^2 t (2 t+u) \right.\nonumber\\&\left. -s (8 t^3+4 t^2 u+7 t u^2+2 u^3) -t (4 t^3-6 t u^2-3 u^3) \right] 
	- Q^2 \left[ s^4 (t+6 u)-2 s^3 (7 t^2+4 t u-u^2) \right.\nonumber\\&\left. -2 s^2 (6 t^3+13 t^2 u+3 t u^2-u^3)  -2 s u (4 t^3+t^2 u-4 t u^2-3 u^3)-t u^2 (4 t^2+6 t u+3 u^2) \right] \nonumber\\&\left.
	+ s (s+u) \left[ s^3 (4 t+3 u)+s^2 (4 t^2+6 t u+u^2)+s u^3+u^2 (4 t^2+6 t u+3 u^2) \right]
	\right\}
	\, ,
	\nonumber \\
	\widehat{w}_L^\perp & = 32 Q^{10}  t^3 
	- 16 Q^8 t^2 \left[ s^2 + 2 s (t - u) + 4 t^2 + 6 t u + u^2 \right] 
	+ 8 Q^6 t \left[ 2 s^3 (3 t - u)  + s^2 t (7 t - 2 u) \right.\nonumber\\&\left.  + 2 s (2 t^3 - 3 t^2 u - 7 t u^2 - u^3) + t (4 t^3 + 12 t^2 u + 11 t u^2 + 2 u^3) \right]
	+ 2 Q^4 t \left[ s^4 (24 u - 7 t) \right.\nonumber\\&\left. - 4 s^3 (4 t^2 + t u - 2 u^2) + 2 s^2 u (16 t^2 + 15 t u - 4 u^2) + 4 s u (8 t^3 + 20 t^2 u + 15 t u^2 + 2 u^3) + t u^4 \right]  \nonumber\\&
	+ 2 Q^2 s t \left[ 5 s^5 + s^4 (8 t - 11 u) + 2 s^3 (4 t^2 + 4 t u - 9 u^2) + 2 s^2 u (8 t^2 + 20 t u + 9 u^2) \right.\nonumber\\&\left. + s u^2 (24 t^2 + 40 t u + 21 u^2) + u^5 \right] 
	+ s^2 (s + u)^2 \left[ s^4 + 2 s^3 u + 2 s^2 u (8 t + u) \right.\nonumber\\&\left. + 2 s u (8 t^2 + 8 t u + u^2) + u^4 \right]
	\,,
	\nonumber \\
	\widehat{w}_L^\myparallel = & 8 Q^2 \left\{
	16 Q^8 t^3
	+8 Q^6 t^2 \left[ s^2-2 s (t-u)-4 t^2-6 t u-u^2 \right] \right.
	-4 Q^4 t \left[ s^4+2 s^3 t \right.\nonumber\\&\left. +s^2 (5 t^2+10 t u+u^2) -2 s (2 t^3-3 t^2 u-7 t u^2-u^3)-t (4 t^3+12 t^2 u+11 t u^2+2 u^3) \right] \nonumber\\&
	+Q^2 \left[ -2 s^6+4 s^5 (t-u) +s^4 (5 t^2+4 t u-4 u^2) +4 s^3 (4 t^3+t^2 u-3 t u^2-u^3)  \right.\nonumber\\&\left.  +2 s^2 (8 t^4+32 t^3 u+21 t^2 u^2-2 t u^3-u^4) +4 s t u (8 t^3+20 t^2 u+15 t u^2+2 u^3)+t^2 u^4 \right] \nonumber\\&
	+s (s+u) \left[ 2 s^5+2 s^4 (t+2 u)+s^3 u (3 t+4 u) +s^2 u (16 t^2+17 t u+4 u^2) \right.\nonumber\\&\left.\left.  +s u (16 t^3+32 t^2 u+17 t u^2+2 u^3)+t u^4 \right]
	\right\}
	\,,
	\nonumber \\
	\widehat{w}^\perp_{\Delta} = & 2 M_\psi \sqrt{t (Q^2 t + s u)}  \left\{-16 Q^8 t^2 + 
	8 Q^6 t \left[ 2 t^2 + 4 t u + u (u - s) \right]  
	+ 2 Q^4 \left[ 2 s^4 + s^3 (2u - 3 t) \right.\right.\nonumber\\&\left. + s^2 (4 t^2 + 3 t u + 2 u^2) + s (4 t^3 + 12 t^2 u + 11 t u^2 + 2 u^3) - t u (4 t^2 + 8 t u + 3 u^2)\right] \nonumber\\&
	- Q^2 s \left[ 7 s^4 + 4 s^3 (t + 2 u) - 2 s^2 u (6 t + u) - 4 s t u (2 t + u) + u^2 (8 t^2 + 12 t u + 3 u^2)  \right] \nonumber\\&\left.
	+ s^2 (s - u) (s + u)^2 (3 s + 4 t + 3 u) \right\}
	\,,
	\nonumber \\
	\widehat {w}^\myparallel_{\Delta} = & - 8 Q^2M_\psi \sqrt{t (Q^2 t + s u)}  \left\{
	16 Q^6 t^2 
	- 8 Q^4 t \left[ 2 t^2 + 4 t u + u (u - s)  \right]  \right.
	- 2 Q^2 \left[2 s^4 + s^3 (2 u - 3 t) \right.\nonumber\\&\left. + s^2 (4 t^2 + 3 t u + 2 u^2) + s (4 t^3 + 12 t^2 u + 11 t u^2 + 2 u^3) - t u (4 t^2 + 8 t u + 3 u^2)  \right] \nonumber\\&\left.
	+ s (s + u) \left[ 3 s^3 + s^2 (u -4 t) + s (u^2 -8 t^2 - 8 t u) + u (8 t^2 + 12 t u + 3 u^2) \right] \right\}
	\,,\nonumber \\ 
	\widehat{w}^\perp_{\Delta\Delta}  = & 4 M_\psi^2 (Q^2 t + s u) \left\{
	-4 Q^6 t^2 
	+ 2 Q^4 t \left[ (s - t)^2  + (t + u)^2 \right] - Q^2 t \left[ s^3 - s^2 u - s u (4 t + u) + u^3 \right]	\right.\nonumber\\&\left.	
	+ s^2 (s+u)^2 (s + 2t + u)
	\right\}
	\,,\nonumber \\ 
	\widehat{w}^\myparallel_{\Delta\Delta} =  & -16 Q^2 M_\psi^2 t (Q^2 t + s u) \left\{
	4 Q^4 t - 
	2 Q^2 (s^2 - 2 s t + 2 t^2 + 2 t u + u^2) \right.\nonumber\\&\left.
	+ (s + u) \left[ s^2 + u^2 - 2 s (2 t + u) \right]
	\right\} \,.
\end{align}

\item  $\gamma^* g \,\to \,c \bar c [^1S_0^{[8]}]\, g$:
\begin{align}
	{F}  = &  { 32\,  e_c^2 \alpha \alpha_s^2  \over s^2 t (s + t)^2 (s + u)^2 (t + u)^2  M_\psi } \,  \langle 0 \vert {\cal O}_8(\,^1 S_0) \vert  0 \rangle, \nonumber \\
	\widehat{w}_T^\perp =  & 2 Q^6 t^3 u^2
	+2 Q^4 s t^2 u \left[ s (t+u)+t^2+t u+2 u^2 \right] 
	+Q^2 s^2 t \left[ s^4+2 s^3 (t+u)+3 s^2 (t+u)^2 \right.\nonumber\\&\left. +2 s (t^3+3 t^2 u+4 t u^2+2 u^3) +t^4 +2 t^3 u+5 t^2 u^2+4 t u^3+3 u^4 \right] \nonumber\\&
	+s^3 u \left[ s^4+2 s^3 (t+u)+3 s^2 (t+u)^2+2 s (t+u)^3+(t^2+t u+u^2)^2 \right]
	\,, \nonumber \\
	\widehat{w}_T^\myparallel  = & 4 Q^2 t \left\{
	2 Q^4 t^2 u^2+2 Q^2 s t u \left[ t^2+t u+2 u^2 + s (t + u) \right] \right.
	+s^2 \left[ s^2 (t+u)^2  \right.\nonumber\\&\left.\left.+2 s (t^3+2 t^2 u+2 t u^2+u^3)+t^4+2 t^3 u+ 3 t^2 u^2 +2 t u^3 +2 u^4\right] 
	\right\}
	\, ,\nonumber \\
	\widehat{w}_L^\perp = & \widehat {w}_T^\perp
	\,,\nonumber \\
	\widehat{w}_L^\myparallel =&  \widehat{w}_T^\myparallel
	\,,\nonumber \\
	\widehat{w}^\perp_{\Delta} = & \widehat {w}^\myparallel_{\Delta}=  \widehat{w}^\perp_{\Delta\Delta} = \widehat{w}^\myparallel_{\Delta\Delta} = 0 \,.
\end{align}

\item  $\gamma^* g \,\to \,c \bar c [^3S_1^{[1]}]\, g$:
\begin{align}
F = & { 256\,  e_c^2 \alpha \alpha_s^2 \over 27 s^2 (s + t)^2 (s + u)^2 (t + u)^2 \left[4Q^2t + (s + u)^2 \right] M_\psi } \, \langle 0 \vert {\cal O}_1(\,^3 S_1) \vert  0 \rangle, \nonumber\\
\widehat{w}_T^\perp = & M_\psi^2 \left\{
4 Q^8 t^4 
+ 4 Q^6 t^3 \left[ 2 s (s+ u) + t (t - u) \right]  
+ 2 Q^4 t^2 \left[ s^4 - s^3 (t - 6 u) + 2 s^2 (t^2 + 2 u^2) \right.\right.\nonumber\\&\left. + 3 s t (t - u) u + t^2 u^2 \right]
+ Q^2 s t \left[ s^4 (2 u - t) + s^3 (3 t^2 + t u + 8 u^2) + s^2 u (8 t^2 + 3 t u + 4 u^2) \right.\nonumber\\&\left. + s t (5 t - 3 u) u^2 + 2 t^2 u^3 \right]
+ s^2 \left[ s^4 (t^2 + u^2) + 2 s^3 u (t^2 + t u + u^2) + s^2 u^2 (4 t^2 + 2 t u + u^2) \right.\nonumber\\&\left. \left. + 2 s t^2 u^3 + t^2 u^4  \right] \right\}
\,,
\nonumber\\
\widehat{w}_T^\myparallel  = &  8 Q^2 M_\psi^2 t (Q^2 t + s u)  \left\{ t^2 \left[  s^2 + 2Q^2 (Q^2 + t) \right] + t u \left[ s (s + t) + 2Q^2 (s - t) \right] + u^2 \left[ s^2 - s t + t^2 \right] \right\}
\, ,\nonumber \\
\widehat{w}_L^\perp = &  2t \left\{
4 Q^{10} t^3  - 
4 Q^8 t^2 \left[ s (t - 2 u) + 2 t (t + u) \right] 
+ 2 Q^6 t \left[ s^3 t + 2 t^2 (t + u)^2 + s t (2 t^2 - 5 t u - 7 u^2) \right.\right.\nonumber\\&\left. + s^2 (3 t^2 - 2 t u + 3 u^2) \right] 
+ 2 Q^4 s \left[ 3 t^2 u (t + u)^2 + s^3 t (t + 2 u) + s^2 (5 t^2 u + u^3) \right.\nonumber\\&\left. - s t (t^3 - 2 t^2 u + t u^2 + 4 u^3) \right] 
+ Q^2 s^2 \left[ s^4 t - 2 s (t - u)^2 u (t + u) + 2 s^3 (t + u)^2 + 3 t u^2 (t + u)^2 \right.\nonumber\\&\left.\left. + s^2 t (t^2 + 2 t u + 6 u^2) \right] 
+ s^3 u (s + t + u)^2 (s^2 + u^2)
\right\}
\,,\nonumber \\
\widehat{w}_L^\myparallel = & 4 Q^2 \left\{ 
8 Q^8 t^4 
+ 8 Q^6 t^3 \left[ s^2 - s (t - 2 u) - 2 t (t + u) \right] \right.
+ 2 Q^4 t^2 \left[ s^4 + 4 t^2 (t + u)^2 + s^3 (6 u - 4 t) \right.\nonumber\\&\left. +  2 s t (2 t^2 - 5 t u - 7 u^2) - s^2 (4 t^2 + 10 t u - 7 u^2) \right] 
+ 2 Q^2 s t \left[ s^4 (u - t) + 6 t^2 u (t + u)^2 \right.\nonumber\\&\left. + s^3 (t^2 - 3 t u + 4 u^2) + s t (2 t^3 + 8 t^2 u - 3 t u^2 - 9 u^3) + s^2 (4 t^3 - 7 t u^2 + 3 u^3)\right] \nonumber\\&
+ s^2 \left[  s^4 (t^2 + u^2)  + 2 s^3 (t^3 + 2 t^2 u + u^3) + s^2 (t^4 + 6 t^3 u + 2 t^2 u^2 - 4 t u^3 + u^4) \right.\nonumber\\&\left.\left. + 2 s t u (t^3 + 3 t^2 u - 2 u^3) + 5 t^2 u^2 (t + u)^2  \right]  \right\}
\,,\nonumber \\
\widehat{w}^\perp_{\Delta} = & M_\psi \sqrt{t (Q^2 t + s u)} \left\{
- 8 Q^8 t^3 
- 4 Q^6 t^2 (s^2 - s t - 2 t^2 + 3 s u - 3 t u) 
+ 2 Q^4 t \left[ s^3 (t - 2 u) \right.\right.\nonumber\\&\left. + 2 s^2 (t - 2 u) u  + s t u (4 t + 7 u) - 2 t^2 u (t + u) \right] 
- Q^2 s \left[ s^4 t + 2 s^3 (t^2 + t u + u^2) \right.\nonumber\\&\left. + s^2 (3 t^3 + 3 t^2 u - t u^2 + 2 u^3) - 2 s t u (t^2 + 2 t u + 3 u^2) + 3 t^2 u^2 (t + u)  \right] \nonumber\\&\left.
+ s^2 (s - u) \left[ s^3 (t - u) - s u^3 + t u^2 (t + u) + s^2 (t^2 - 2 u^2) \right]
\right\}
\,,\nonumber \\
\widehat {w}^\myparallel_{\Delta} = & - 4 Q^2 M_\psi \sqrt{t (Q^2 t + s u)} \left\{ 
8 Q^6 t^3 
+ 4 Q^4 t^2 \left[ s(s - t + 3 u) - t (2t + 3u) \right]
- 2 Q^2 t \left[s^3 (t - 2 u) \right.\right.\nonumber\\&\left. + 2 s^2 u (t - 2 u) + s t u (4 t + 7 u) - 2 t^2 u (t + u) \right]
+ s \left[ s^3 (t^2 + t u + 2 u^2) + s^2 (t^3 + 3 t^2 u + 2 u^3) \right.\nonumber\\&\left.  \left. - s t u^2 (3 t + 5 u) + 3 t^2 u^2 (t + u)   \right] 
\right\}
\,,\nonumber \\ 
\widehat{w}^\perp_{\Delta\Delta}  = &  M_\psi^2 t (Q^2 t + s u) \left\{
- 4 Q^6 t^2 
- 4 Q^4 t \left[ s u - t (t + u) \right] 
+ 2 Q^2 s \left[ s^2 t - s (t^2 + u^2) + t u (t + u) \right] \right.\nonumber\\&\left.
+ s^2 (s + t + u) (s^2 + u^2)
\right\}
\,,\nonumber \\ 
\widehat{w}^\myparallel_{\Delta\Delta} = &  - 8 Q^2 M_\psi^2 t (Q^2 t + s u) \left[
2 Q^4 t^2 - 2 Q^2 t (t^2 - s u + t u) + s (s - t) u (t + u) 
\right]
\,.
\end{align}

\item  $\gamma^* g \,\to \,c \bar c [^3S_1^{[8]}]\, g$:
\begin{align}
\widehat{w}_\Lambda^{\cal P}[\,^3 S_1^{[8]}] = &  {15 \over 8} \widehat{w}_\Lambda^{\cal P}[\,^3 S_1^{[1]}] \,.
\end{align}

\item  $\gamma^* g \,\to \,c \bar c [^3P_J^{[8]}]\, g$:

The formulae for the $P$-wave components of $\widehat {w}^{\cal P}_T$ and $\widehat w^{\cal P}_L$  are too lengthy to be shown here. They are available upon request as a Wolfram Mathematica notebook file. Only $\widehat {w}^{\cal P}_\Delta$ and $\widehat{w}^{\cal P}_{\Delta\Delta}$ are given below in an analytical form. 
\begin{align}
F =  & -  { 192\, \sqrt{t (Q^2 t + s u)} \, e_c^2 \alpha \alpha_s^2  \over s^2 t^2 (s + t)^3 (s + u)^4 (t + u)^3 \left[4Q^2t + (s + u)^2\right] M_\psi^2  } \, \langle 0 \vert {\cal O}_8(\,^3 P_0) \vert  0 \rangle, \nonumber
\end{align}
\begin{align}
\widehat{w}^\perp_{\Delta} = & t\left\{
64 Q^{10} t^3 (s^3 t - s t^2 u - s^2 u^2 + t u^3) 
- 16 Q^8 t^2 \left[2 s^5 u + s^4 (6 t^2 - 2 t u + 6 u^2)  \right.\right.\nonumber\\& + s^3 (5 t^3 + 11 t^2 u + 10 u^3)
- s^2 (t^4 + 7 t^3 u + 6 t u^3 - 4 u^4) - t u^2 (t^3 - 5 t^2 u - 10 t u^2 - 2 u^3) \nonumber\\&\left. - s u (6 t^4 + 11 t^3 u - 5 t^2 u^2 + 2 t u^3 - 2 u^4) \right]
- 8 Q^6 t \left[2 s^7 (t + u) + 2 s^6 (t^2 + t u + 5 u^2)  \right.\nonumber\\& - s^5 (5 t^3 - 6 t^2 u + 6 t u^2 - 20 u^3)
- s^4 (8 t^4 + 8 t^3 u - 18 t^2 u^2 + 5 t u^3 - 24 u^4) \nonumber\\& - s^3 (t^5 + 9 t^4 u + 14 t^3 u^2 - 19 t^2 u^3 + 9 t u^4 - 14 u^5) + t^2 u^2 (2 t^4 + 3 t^3 u - 7 t^2 u^2 - 12 t u^3 - 3 u^4) \nonumber\\&
+ s^2 (2 t^6 + 15 t^5 u + 23 t^4 u^2 + 16 t^3 u^3 + 43 t^2 u^4 + 5 t u^5 + 6 u^6) \nonumber\\&\left.
+ s t u (4 t^5 + 23 t^4 u + 29 t^3 u^2 + 23 t^2 u^3 + 27 t u^4 + 3 u^5) \right] 
+ 4 Q^4 \left[ s^8 (6 t^2 - 4 u^2) \right.\nonumber\\& + s^7 (15 t^3 + 23 t^2 u + 12 t u^2 - 12 u^3) + s^6 (17 t^4 + 37 t^3 u + 32 t^2 u^2 + 32 t u^3 - 20 u^4) \nonumber\\&
+ s^5 (6 t^5 + 28 t^4 u + 40 t^3 u^2 + 21 t^2 u^3 + 39 t u^4 - 20 u^5)  \nonumber\\& 
+ s^4 (4 t^6 + 10 t^5 u + 33 t^4 u^2 + 56 t^3 u^3 + 6 t^2 u^4 + 31 t u^5 - 12 u^6)\nonumber\\&
+ s^3 (2 t^7 + 6 t^6 u - 3 t^5 u^2 - 4 t^4 u^3 + 17 t^3 u^4 - 25 t^2 u^5 + 9 t u^6 - 4 u^7) \nonumber\\&
+ s^2 t u (6 t^6 + 20 t^5 u + 13 t^4 u^2 + 11 t^3 u^3 + 31 t^2 u^4 + 5 u^6)   + t^4 u^3 (2 t^3 + 4 t^2 u + t u^2 - u^3)  
\nonumber
\\&\left.
+ s t^2 u^2 (6 t^5 + 22 t^4 u + 37 t^3 u^2 + 52 t^2 u^3 + 44 t u^4 + 9 u^5)\right] 
+ 2 Q^2 s \left[ s^8 (9 u^2 + 12 t u - 5 t^2)  \right.\nonumber\\& 
- s^7 (13 t^3 - 9 t^2 u - 43 t u^2 - 29 u^3) - s^6 (19 t^4 - 2 t^3 u - 35 t^2 u^2 - 64 t u^3 - 52 u^4)\nonumber\\& 
- s^5 (9 t^5 - t^4 u - 53 t^3 u^2 - 83 t^2 u^3 - 88 t u^4 - 62 u^5) \nonumber\\&
+ s^4 u (16 t^5 + 60 t^4 u + 110 t^3 u^2 + 145 t^2 u^3 + 80 t u^4 + 45 u^5) \nonumber\\&
+ s^3 u (6 t^6 + 34 t^5 u + 68 t^4 u^2 + 105 t^3 u^3 + 127 t^2 u^4 + 47 t u^5 + 21 u^6) \nonumber\\&
+ s^2 u^2 (14 t^6 + 52 t^5 u + 95 t^4 u^2 + 114 t^3 u^3 + 105 t^2 u^4 + 28 t u^5 + 6 u^6) \nonumber\\&\left.
+ s t u^3 (10 t^5 + 47 t^4 u + 83 t^3 u^2 + 79 t^2 u^3 + 45 t u^4 + 6 u^5) + 2 t^3 u^4 (t^3 + 2 t^2 u - u^3) \right]  \nonumber\\&
- s^2 (s^2 - u^2) \left[s^6 (3 u^2 + 10 t u - t^2)  + s^5 (t^3 + 28 t^2 u + 33 t u^2 + 6 u^3) \right.\nonumber\\&
+ s^4 (3 t^4 + 45 t^3 u + 49 t^2 u^2 + 41 t u^3 + 6 u^4) + s^3 (t^5 + 30 t^4 u + 46 t^3 u^2 + 40 t^2 u^3 + 41 t u^4 + 6 u^5)
\nonumber\\& + s^2 u (7 t^5 + 22 t^4 u + 46 t^3 u^2 + 49 t^2 u^3 + 33 t u^4 + 3 u^5) \nonumber\\&\left.\left. + s t u^2 (7 t^4 + 30 t^3 u + 45 t^2 u^2 + 28 t u^3 + 10 u^4) 
+ t^2 u^3 (t^3 + 3 t^2 u + t u^2 - u^3)\right] 
\right\}
\,,
\nonumber 
\\
\widehat {w}^\myparallel_{\Delta} = & 8 Q^2 t \left\{
32 Q^8 t^3 (s^3 t - s t^2 u - s^2 u^2 + t u^3) 
- 8 Q^6 t^2 \left[ 2 s^5 u + s^4 (6 t^2 - 2 t u + 6 u^2)  \right.\right.
\nonumber
\\&  + s^3 (5 t^3 + 11 t^2 u + 10 u^3) - s^2 (t^4 + 7 t^3 u + 6 t u^3 - 4 u^4) \nonumber\\&\left. 
- s u (6 t^4 + 11 t^3 u - 5 t^2 u^2 + 2 t u^3 - 2 u^4) +  t u^2 (2 u^3 + 10 t u^2 + 5 t^2 u - t^3)  \right] \nonumber\\&
- 4 Q^4 t \left[ 2 s^7 (t + u) + 2 s^6 (2 t^2 + 2 t u + 5 u^2)  + s^5 (20 u^3 - 4 t u^2 + 10 t^2 u - 5 t^3) \right.\nonumber\\& 
- s^4 (8 t^4 + 8 t^3 u - 18 t^2 u^2 + 7 t u^3 - 24 u^4)  - s^3 (t^5 + 9 t^4 u + 14 t^3 u^2 - 15 t^2 u^3 + 11 t u^4 - 14 u^5) \nonumber\\&
+ s^2 (2 t^6 + 15 t^5 u + 23 t^4 u^2 + 16 t^3 u^3 + 41 t^2 u^4 + 5 t u^5 + 6 u^6) \nonumber\\&
+ s t u (4 t^5 + 23 t^4 u + 29 t^3 u^2 + 23 t^2 u^3 + 27 t u^4 + 3 u^5) \nonumber\\&\left.  + t^2 u^2 (2 t^4 + 3 t^3 u - 7 t^2 u^2 - 12 t u^3 - 3 u^4) \right]
+ 2 Q^2 \left[ 2 s^8 (t^2 - t u - 2 u^2) \right.\nonumber\\& + s^7 (7 t^3 + 9 t^2 u + 6 t u^2 - 12 u^3) 
+ s^6 (3 t^4 + 27 t^3 u + 20 t^2 u^2 + 28 t u^3 - 20 u^4) \nonumber\\& + s^5 u (16 t^4 + 42 t^3 u + 21 t^2 u^2 + 43 t u^3 - 20 u^4) \nonumber\\&
+ s^4 (4 t^6 + 8 t^5 u + 33 t^4 u^2 + 54 t^3 u^3 + 18 t^2 u^4 + 37 t u^5 - 12 u^6) \nonumber\\&
+ s^3 (2 t^7 + 6 t^6 u - t^5 u^2 + 8 t^4 u^3 + 27 t^3 u^4 -  11 t^2 u^5 + 11 t u^6 - 4 u^7) \nonumber\\& 
+ s^2 t u (6 t^6 + 20 t^5 u + 19 t^4 u^2 + 25 t^3 u^3 + 39 t^2 u^4 + 4 t u^5 + 5 u^6) + t^4 u^3 (2 t^3 + 4 t^2 u + t u^2 - u^3)  \nonumber\\&\left.
+ s t^2 u^2 (6 t^5 + 22 t^4 u + 37 t^3 u^2 + 52 t^2 u^3 + 44 t u^4 + 9 u^5) 
\right] 
+ s (s + u) \left[ s^7 (3 t^2 + 10 t u + 7 u^2) \right.\nonumber\\& + s^6 (8 t^3 + 26 t^2 u + 30 t u^2 + 20 u^3) 
+ s^5 (7 t^4 + 20 t^3 u + 43 t^2 u^2 + 42 t u^3 + 32 u^4)  \nonumber\\&
+ 2 s^4 (t^5 + 9 t^4 u + 27 t^3 u^2 + 35 t^2 u^3 + 23 t u^4 + 15 u^5) \nonumber\\&
+ s^3 u (20 t^5 + 48 t^4 u + 56 t^3 u^2 + 45 t^2 u^3 + 26 t u^4 +  17 u^5)  \nonumber\\&
+ 2 s^2 u (3 t^6 + 7 t^5 u + 7 t^4 u^2 + 14 t^3 u^3 + 24 t^2 u^4 +  12 t u^5 + 3 u^6) \nonumber\\&\left.\left.
+ s t u^2 (8 t^5 + 32 t^4 u + 57 t^3 u^2 + 60 t^2 u^3 + 37 t u^4 + 6 u^5)  + 2 t^3 u^3 (t^3 + 2 t^2 u - u^3)\right] 
\right\}
\,,
\nonumber 
\end{align}
\begin{align}
\widehat{w}^\perp_{\Delta\Delta}  =  & 2 M_\psi \sqrt{t (Q^2 t + s u)} \left\{
16 Q^8 t^3 (s^3 t - s^2 u^2 - s t^2 u + t u^3) 
- 4 Q^6 t^2 \left[ 2 s^5 (t + u) + 4 s^4 (t^2 + u^2) \right.\right.
\nonumber
\\& + s^3 (5 t^3 + 9 t^2 u + 2 t u^2 + 8 u^3)   - s^2 (t^4 + 3 t^3 u + 4 t u^3 - 4 u^4)  - t u^2 (t^3 - 5 t^2 u - 6 t u^2 - 2 u^3)  \nonumber\\&\left. 
- s u (6 t^4 + 7 t^3 u - t^2 u^2 + 2 t u^3 - 2 u^4)\right] 
- 4 Q^4 t \left[ s^6 (t^2 + t u + 2 u^2) \right.\nonumber\\&  - s^5 (2 t^3 - 3 t^2 u + 2 t u^2 - 4 u^3) - s^4 (t^4 - t^3 u - 8 t^2 u^2 + 2 t u^3 - 6 u^4)   \nonumber\\& 
+ s^3 u (t^4 + t^3 u + 6 t^2 u^2 - 3 t u^3 + 4 u^4) + s^2 (t^6 + 8 t^5 u + 9 t^4 u^2 + 8 t^3 u^3 + 10 t^2 u^4 + t u^5 + 2 u^6) \nonumber\\&\left.
+ s t u (2 t^5 + 10 t^4 u + 11 t^3 u^2 + 13 t^2 u^3 + 7 t u^4 + u^5) 
+ t^2 u^2 (t^4 + 2 t^3 u - t u^3 - u^4) \right] \nonumber\\& 
- 2 Q^2 s t \left[ 2 s^7 t + s^6 (3 t^2 + 3 t u - 5 u^2)  + 2 s^5 (2 t^3 - t^2 u - t u^2 - 7 u^3) \right.\nonumber\\& 
+ s^4 (5 t^4 + 5 t^3 u - 12 t^2 u^2 - 6 t u^3 - 18 u^4) + s^3 (5 t^4 u - 16 t^2 u^3 - 5 t u^4 - 12 u^5) \nonumber\\&
+ s^2 u^2 (11 t^4 + 9 t^3 u + t^2 u^2 + 3 t u^3 - 5 u^4) +s u^3 (11 t^4 + 10 t^3 u + 6 t^2 u^2 - t u^3 - 2 u^4) \nonumber\\&\left. - 2 t u^5 (2 t^2 + 2 t u + u^2)  \right] 
- s^2 (s + u) \left[ s^6 t (u - t) - 2 s^5 (t - u)^2 (t + u)  \right.\nonumber\\& + s^4 (t^4 + 9 t^3 u - 3 t^2 u^2 - 3 t u^3 - 4 u^4) 
+ s^3 u (12 t^4 + 3 t^3 u - 12 t^2 u^2 - 3 t u^3 - 2 u^4) \nonumber\\& 
+ s^2 t u (2 t^4 + 6 t^3 u + 3 t^2 u^2 - 3 t u^3 + 2 u^4) 
+ s t u^2 (2 t^4 + 12 t^3 u + 9 t^2 u^2 + 2 t u^3 + u^4) \nonumber\\&\left.\left. + t^2 u^4 (t^2 - 2 t u - u^2)  \right]
\right\}
\,,
\nonumber\\
\widehat{w}^\myparallel_{\Delta\Delta} =  & 16 Q^2  M_\psi t \sqrt{t (Q^2 t + s u)} \left\{
8 Q^6 t^2 (s^3 t- s^2 u^2  - s t^2 u + t u^3) 
- 2 Q^4 t \left[ 2 s^5 (t + u) + 4 s^4 (t^2 + u^2)  \right.\right.\nonumber\\& + s^3 (5 t^3 + 9 t^2 u + 2 t u^2 + 8 u^3)  - s^2 (t^4 + 3 t^3 u + 4 t u^3 - 4 u^4) \nonumber\\&\left. 
- s u (6 t^4 + 7 t^3 u - t^2 u^2 + 2 t u^3 - 2 u^4) - t u^2 (t^3 - 5 t^2 u - 6 t u^2 - 2 u^3) \right] \nonumber\\&
- 2 Q^2 \left[ 2 s^6 u (t + u) + s^5 u (t^2 + t u + 4 u^2) - s^4 (t^4 - t^3 u - 4 t^2 u^2 - t u^3 - 6 u^4) \right.\nonumber\\&
- s^3 u (t^4 - t^3 u - 4 t^2 u^2 + 2 t u^3 - 4 u^4)  + s^2 (t^6 + 8 t^5 u + 9 t^4 u^2 + 10 t^3 u^3 + 9 t^2 u^4 + t u^5 + 2 u^6) \nonumber\\&\left.
+ s t u (2 t^5 + 10 t^4 u + 11 t^3 u^2 + 13 t^2 u^3 + 7 t u^4 + u^5) + t^2 u^2 (t^4 + 2 t^3 u - t u^3 - u^4) \right] \nonumber\\&
+ s (s + u) \left[ 3 s^5 (t + u)^2 +  s^4 (4 t^3 + 9 t^2 u + 10 t u^2 + 5 u^3) +  s^3 u (6 t^3 + 15 t^2 u + 14 t u^2 + 7 u^3) \right.\nonumber\\& - 
s^2 (t^5 - 2 t^4 u - 12 t^3 u^2 - 13 t^2 u^3 - 5 t u^4 - 3 u^5) \nonumber\\&\left. \left.
- s u (t^5 + 6 t^4 u + 6 t^3 u^2 + 4 t^2 u^3 - t u^4 - 2 u^5) + 2 t u^4 (2 t^2 + 2 t u + u^2) \right] 
\right\}
\,.
\end{align}

\end{itemize}

\end{document}